\documentclass[11pt,a4paper]{article} 
\pdfoutput=1
\usepackage{jheppub}
\usepackage[T1]{fontenc}
\usepackage{mathbbol}
\usepackage{mathrsfs}

\usepackage{bm}
\usepackage{booktabs}
\usepackage{colortbl}
\usepackage{enumerate}
\usepackage{latexsym}
\usepackage{lscape}
\usepackage{multirow}
\usepackage{psfrag}

\def\beq{\begin{equation}}
\def\eeq{\end{equation}}
\def\beqa{\begin{eqnarray}}
\def\eeqa{\end{eqnarray}}

\def\be{\begin{equation}}
\def\ee{\end{equation}}
\def\bea{\begin{eqnarray}}
\def\eea{\end{eqnarray}}

\def\nn{\nonumber}

\newcommand\Eqn[1]     {Eq.~\eqref{#1}}

\newcommand\eqn[1]     {eq.~\eqref{#1}}
\newcommand\eqns[2]    {eqs.~\eqref{#1} and~\eqref{#2}}

\newcommand{\bei}{\begin{itemize}}
\newcommand{\eei}{\end{itemize}}

\newcommand{\RE}{{\rm Re}}

\def\T{{\bf T}}
\def\Tt{{\bf T}_t^2} % square of Casimir
\def\Tsu{{\bf T}_{s{-}u}^2}

\newcommand{\bzero}{B_0} % (\eps)
\newcommand{\bn}[1]{B_{#1}} % (\eps)
\newcommand{\hbn}[1]{\hat B_{#1}} % (\eps)
\newcommand{\Dd}{\mathrm{d}}
\newcommand{\DD}{\mathrm{D}}
\newcommand{\Dk}{\DD k}
\newcommand{\dk}{\Dd k}
\newcommand{\LL}{{\mathrm{LL}}}
\newcommand{\NLL}{{\mathrm{NLL}}}
\newcommand{\NNLL}{{\mathrm{NNLL}}}

\newcommand{\Wt}[1]{{\Omega}^{(#1)}}
\newcommand{\Wts}[1]{{\Omega}_s^{(#1)}}
\newcommand{\mratio}[2]{\left(\frac{#1^2}{#2^2}\right)}

\newcommand{\Mbar}{\bar{\mathcal{M}}}
\newcommand{\R}{R}
\newcommand{\po}{\mathcal P}

\newcommand{\CA}{C_{A}}

\newcommand{\nf}{n_f}

\newcommand{\as}{\alpha_s}

\newcommand{\eps}{\epsilon}

\newcommand{\ord}{{\cal O}}

\def\bra#1{\langle#1|}
\def\ket#1{|#1\rangle}

\newcommand{\LO}{{\rm(LO)}}
\newcommand{\NLO}{{\rm(NLO)}}

\def\MM{\mathcal{M}}
\def\Mreduced{\hat{\mathcal{M}}}
\def\Mtree{\mathcal{M}^{(\rm tree)}}

\def\Hhard{\mathcal{H}} % hard function
\def\Hhat{\hat H} % subtracted Hamiltonian 

\def\Z{{\bf Z}} % renomalisation factor

\def\Log{L} % =\log(|s/t|)-i\pi/2

\allowdisplaybreaks

\title{Infrared singularities of QCD scattering amplitudes in the Regge limit to all orders}

\author[a]{Simon Caron-Huot,}
\author[b]{Einan Gardi,}
\author[b]{Joscha Reichel,}
\author[b,c]{Leonardo Vernazza}

\affiliation[a]{Department of Physics, McGill University, 3600 rue University, Montr\'eal, QC Canada H3A 2T8}
\affiliation[b]{Higgs Centre for Theoretical Physics, 
School of Physics and Astronomy, 
The University of Edinburgh, Edinburgh EH9 3FD, Scotland, UK}
\affiliation[c]{Nikhef, Science Park 105, NL--1098 XG Amsterdam, The Netherlands}

\emailAdd{schuot@physics.mcgill.ca}
\emailAdd{Einan.Gardi@ed.ac.uk}
\emailAdd{joscha.reichel@ed.ac.uk}
\emailAdd{l.vernazza@nikhef.nl}

\abstract{Scattering amplitudes of partons in QCD 
contain infrared divergences which can be resummed 
to all orders in terms of an anomalous dimension. 
Independently, in the limit of high-energy forward 
scattering, large logarithms of the energy can be 
resummed using Balitsky-Fadin-Kuraev-Lipatov 
theory. We use the latter to analyze the infrared-singular 
part of amplitudes to all orders in perturbation theory 
and to next-to-leading-logarithm accuracy in the 
high-energy limit, resumming the two-Reggeon 
contribution. Remarkably, we find a closed form 
for the infrared-singular part, predicting the Regge 
limit of the soft anomalous dimension to any loop 
order.}


\dedicated{In memory of Lev Nikolaevich Lipatov and his pioneering contributions}

\begin{document}

\begin{flushright}
Edinburgh 2017/24\\
NIKHEF/2017-060\\
\vspace*{-25pt}
\end{flushright}

\maketitle
\flushbottom

%%%%%%%%%%%%%%%%%%%%%%%%%%%%%%%%%%%%%%%%%

\section{Introduction}
\label{intro}

The high-energy limit of QCD scattering has always 
been a subject of much theoretical interest, see 
e.g.~\cite{Kuraev:1977fs,Balitsky:1978ic,Lipatov:1985uk,Mueller:1993rr,Mueller:1994jq,Brower:2006ea,Moult:2017xpp}. 
In particular, the Balitsky-Fadin-Kuraev-Lipatov (BFKL) 
equation~\cite{Kuraev:1977fs,Balitsky:1978ic} provides 
a theoretical framework to resum high-energy (or rapidity) 
logarithms to all orders in perturbation theory. It was used 
extensively to investigate a range of physical phenomena 
including the small-$x$ behaviour of deep-inelastic structure 
functions and parton densities, and jet production with large 
rapidity gaps. The non-linear generalisations of BFKL, known 
as the Balitsky-JIMWLK equation~\cite{Balitsky:1995ub,Balitsky:1998kc,Kovchegov:1999yj,JalilianMarian:1996xn,JalilianMarian:1997gr,Iancu:2001ad}, 
extends the range of phenomena further, e.g.\ 
to describe gluon saturation in heavy-ion collisions.

On the theoretical front, a separate line of investigation 
concerns the structure of partonic scattering amplitudes 
in the high-energy limit~\cite{Sotiropoulos:1993rd,Korchemsky:1993hr,Korchemskaya:1996je,Korchemskaya:1994qp,DelDuca:2001gu,DelDuca:2013ara,DelDuca:2014cya,Bret:2011xm,DelDuca:2011ae,Caron-Huot:2013fea,Caron-Huot:2017fxr}. 
Scattering amplitudes of quarks and gluons are dominated 
at high energies by the $t$-channel exchange of effective 
excitations dubbed Reggeized gluons.  In this context the 
BFKL equation and its generalisations provide again a 
highly-valuable tool: by solving these equations iteratively 
one can compute high-energy logarithms order-by-order 
in perturbation theory \cite{Caron-Huot:2013fea,Caron-Huot:2017fxr}.

The real part of a $2\to 2$ partonic amplitude (i.e.\ its 
signature-odd part, see \eqn{Odd-Even-Amp-Def}) is 
governed by an odd number of Reggeized gluons. The 
leading high-energy logarithms simply exponentiate, 
dressing the $t$-channel gluon propagator by a power of $s/t$.
In Regge theory (see e.g.~\cite{Collins:1977jy}) this 
behaviour corresponds to a Regge pole in the complex 
angular momentum plane. QCD amplitudes can thus 
be factorized in the high-energy limit into a $t$-channel 
Reggeized gluon exchange which captures the 
dependence on the energy, and energy-independent 
impact factors that depend on the colliding partons.
However, this simple picture does not extend beyond 
next-to-leading logarithms (NLL) due to multiple 
Reggeized gluon exchange, which form Regge cuts. 
This was recently demonstrated explicitly
in ref.~\cite{Caron-Huot:2017fxr}, where these effects 
were computed through three-loops, by 
constructing an iterative solution of the relevant BFKL 
or Balitsky-JIMWLK equation, describing the evolution 
of three Reggeized gluons and their mixing with a 
single Reggeized gluon.

In the this paper we extend this study, focusing on the 
imaginary part of $2\to 2$ partonic amplitudes, which 
are governed by the exchange of an even number of 
Reggeized gluons, which also form Regge cuts. The 
leading logarithmic corrections to the even amplitude 
are determined to all orders by a wavefunction of a 
pair of Reggeized gluons, which solves the celebrated 
BFKL evolution equation. This iterative solution, which 
will be central to the present work, can be famously 
described by ladder graphs, where an additional rung 
is generated at each order in the loop expansion. 
 
The study of scattering amplitudes in the high-energy 
limit~\cite{Sotiropoulos:1993rd,Korchemsky:1993hr,Korchemskaya:1996je,Korchemskaya:1994qp,DelDuca:2001gu,DelDuca:2013ara,DelDuca:2014cya,Bret:2011xm,DelDuca:2011ae,Caron-Huot:2013fea,Caron-Huot:2017fxr} 
is intimately linked to the study of their infrared 
singularity structure. Indeed, the gluon Regge 
trajectory $\alpha_g(t)$ is infrared-singular, and 
its exponentiation along with the energy logarithms, 
which is a manifestation of Reggeization, is readily 
consistent with the exponentiation of soft singularities 
through the relevant renormalization group equation. 
The latter of course holds also away from the high-energy 
limit, as guaranteed by infrared factorization theorems.
The correspondence between the structure of amplitudes 
in the high-energy limit, which is governed by rapidity 
evolution equations, on the one hand, and the structure 
of infrared singularities on the other, becomes more 
complicated at subleading orders. While both separately 
provide means to explore the structure of amplitudes 
to all orders in perturbation theory, the interplay 
between the two provides additional insight in either 
direction, as demonstrated multiple times over the 
past few years~\cite{DelDuca:2013ara,DelDuca:2014cya,Bret:2011xm,DelDuca:2011ae,Caron-Huot:2013fea,Caron-Huot:2017fxr}.  

Infrared singularities of massless scattering amplitudes 
are now fully known, for general colour, kinematics and 
any number of partons, through three loops, owing to 
an explicit computation of the soft anomalous dimension 
at this order~\cite{Almelid:2015jia,Gardi:2016ttq}. While 
through two loops infrared singularities are governed 
exclusively by a sum over colour dipoles formed by 
pairs of the hard-scattered partons~\cite{Aybat:2006mz,Gardi:2009qi,Becher:2009cu,Becher:2009qa}, 
at three loops one encounters for the first time infrared 
singularities that are simultaneously sensitive to the 
colour and kinematics of three and four hard partons.
Subsequently, ref.~\cite{Caron-Huot:2017fxr} specialised 
these results to the high-energy limit, and provided a 
detailed comparison between the singularity structure 
deduced from the soft anomalous dimension and what 
has been established there through three loops via 
computations in the high-energy limit. While full 
consistency was found, remarkably, it was shown 
that at three loops (see eq.~(4.11) there) the real 
part of the amplitude is only sensitive to non-dipole 
corrections starting at N$^3$LL accuracy, while for 
the imaginary part of the amplitude they appear 
already at NNLL accuracy. 

As an application of the interplay between these limits,
it was recently demonstrated~\cite{Almelid:2017qju} 
that the functional form of the three-loop soft anomalous 
dimension in general kinematics can in fact be fully 
recovered via a bootstrap procedure using the 
high-energy limit of $2\to 2$ scattering, alongside 
other information, as input. The bootstrap programme 
of the soft anomalous dimension can be extended 
beyond three loops, provided that information from 
special kinematic limits is available. The imaginary 
part of $2\to2$ amplitudes is a natural place to start; 
indeed, already in ref.~\cite{Caron-Huot:2013fea}, 
a non-dipole contribution at four-loops and NLL 
accuracy could be predicted using BFKL theory.

In the present paper we continue to develop this 
line of investigation of the high-energy limit of 
$2\to 2$ scattering, focusing on the imaginary 
(signature-even) part of the amplitude, which is 
governed, as mentioned above, by the exchange 
of a pair of Reggeized gluons satisfying the BFKL 
evolution equation. The leading-order equation is 
sufficient to determine an infinite tower of high-energy 
logarithms in the soft anomalous dimension\footnote{We 
refer to these as next-to-leading logarithms, owing to 
their suppression by one logarithm compared to the 
Reggized-gluon corrections to the real part of the amplitudes.}.

Although the BFKL Hamiltonian has been diagonalised 
in many instances \cite{Lipatov:1985uk}, to study partonic 
amplitudes requires us to use the dimensionally-regulated 
Hamiltonian, which is comparatively less understood.
We will nonetheless find an exact iterative solution! This 
hinges on the following reasons: the two-Reggeon 
wavefunction itself turns out to be finite at all orders, 
so that infrared divergences are controlled by the limit 
of the wavefunction where a Reggeized gluon becomes 
soft.  The evolution equation then closes within that limit, 
dramatically simplifying its solution.
This will enable us to obtain the soft limit of the 
two-Reggeon wavefunction to all loop orders 
and NLL accuracy, and corresponding closed-form
expressions for the singular part of the amplitude 
(see \eqn{eq:mredregge}) and soft anomalous 
dimension (see \eqn{G_NLL_Gl} with (\ref{Gl})), 
which turns out to be an entire function of the 
coupling.

The structure of the paper is as follows. In section 
\ref{dells} we recall the basic notions regarding the 
high-energy limit of $2\to2$ amplitudes and explain 
how the BFKL evolution equation can be solved 
iteratively to determine the two Reggeized gluon 
wavefunction and the imaginary part of the amplitude. 
In section \ref{dells} we also reformulate the equation 
so as to explicitly display the fact that the evolution 
retains infared-finiteness, comment on the symmetries 
displayed by the evolution and recover the four-loop 
results of ref.~\cite{Caron-Huot:2013fea}.  
Appendix \ref{Review} completes this review by 
explaining how the particular form of the evolution 
equation used here follows from the more general 
non-linear set up used in refs.~\cite{Balitsky:1995ub,Caron-Huot:2013fea,Caron-Huot:2017fxr}.
In section \ref{soft} we consider the soft approximation, 
show that the evolution closes in this limit, and exploit 
this simplification to derive all-order solutions for the 
wavefunction and  amplitude.  Finally in section \ref{IRfact} 
we study the implications of our results in the high-energy 
limit regarding the soft anomalous dimension, obtaining 
a closed-form solution for the latter at next-to-leading 
order in high-energy logarithms to all orders, and 
verify the consistency of our BFKL-based result 
with infrared exponentiation. 
 
\section{Scattering amplitudes by iterated solution of the BFKL equation}\label{dells}

The well-known BFKL evolution equation predicts 
the rapidity dependence of two-parton amplitudes 
in the high-energy limit \cite{Kuraev:1977fs,Balitsky:1978ic}.
In the following we briefly summarise the 
conclusions from this approach regarding 
the leading contributions to the signature-even 
amplitude, or the two-Reggeon cut. 

\subsection{The even amplitude from the BFKL wavefunction}
\label{NLL-BFKL-general}

\begin{figure}[htb]
\begin{center}
  \includegraphics[width=0.34\textwidth]{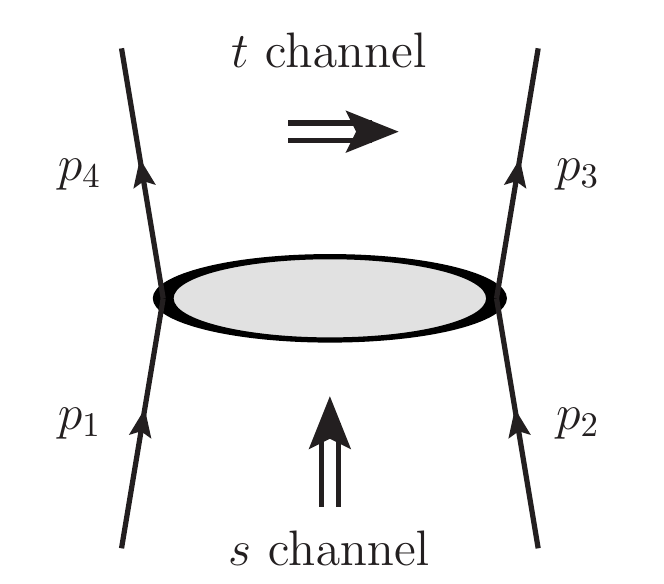}
  \end{center}
  \caption{The $t$-channel exchange dominating the high-energy limit, 
  $s\gg -t>0$. The figure also defines our conventions for momenta assignment 
  and Mandelstam invariants. We shall assume that particles 2 and 3 (1 and 4) are 
  of the same type and have the same helicity.}
\label{setup_fig}
\end{figure}
Let us consider a $2\to 2$ scattering amplitude 
${\cal M}_{ij\to ij}$, where $i,j$ can be a quark 
or a gluon. The momenta are assigned as 
indicated in figure \ref{setup_fig}. In the following 
we will suppress the species indices $i,j$, unless 
explicitly needed. The high-energy limit corresponds 
to a configuration of forward scattering, such that the 
Mandelstam variables satisfy $s\gg -t>0$. 
In analysing this limit it is convenient to 
decompose the amplitude into its odd and even 
components with respect to $s \leftrightarrow u$ 
exchange, the so-called \emph{signature}:
\begin{equation}\label{Odd-Even-Amp-Def}
 {\cal M}^{(\pm)}(s,t) = \tfrac12\Big( {\cal M}(s,t) \pm {\cal M}(-s-t,t) \Big),\\
\end{equation}
where ${\cal M}^{(+)}$, ${\cal M}^{(-)}$
are referred to, respectively, as the \emph{even}
and \emph{odd} amplitudes. As shown 
in ref.~\cite{Caron-Huot:2017fxr}, these have respectively 
\emph{real} and \emph{imaginary} coefficients, when expressed 
in terms of the natural signature-even combination of 
logarithms,
\begin{equation}
\label{L-def}
\frac12\left(\log\frac{-s-i0}{-t}+\log\frac{-u-i0}{-t}\right)
\simeq \log\left|\frac{s}{t}\right| -i\frac{\pi}{2} \equiv L,
\end{equation}
and have independent factorisation properties 
in the high-energy limit. The effect we discuss 
in the following originates from the exchange of
two Reggeons, therefore it proves useful\footnote{The 
full advantage of considering the reduced amplitude 
will become clear in what follows. First,  BFKL evolution 
of the reduced amplitude involves an extra term 
proportional to $\T_t^2$ in (\ref{Jp-def1}). This term 
renders the wavefunction finite. Second, upon 
performing infrared factorization of the reduced 
amplitude one is able to identify the NLL terms 
that originate in the soft anomalous dimension --- 
see eq.~(\ref{Mreduced-IR-1}).} to define a 
\emph{reduced amplitude}, as introduced in 
ref.~\cite{Caron-Huot:2017fxr}, dividing by the 
effect of one-Reggeon exchange:
\be\label{Mreduced}
\Mreduced_{ij\to ij} \equiv 
\,e^{-\,\T_t^2\, \alpha_g(t) \, L} \, \MM_{ij\to ij}\,,
\ee
where $\T_t^2$ represents the total colour charge 
 exchanged in the $t$ channel (see \eqn{TtTsTu} below).
The function $\alpha_g(t)$ in \eqn{Mreduced} represents 
the gluon Regge trajectory having the perturbative 
expansion
\be\label{GluonRegge}
\alpha_g(t)  \,= \,
\sum_{n = 1}^{\infty} \left( \frac{\as}{\pi} \right)^n 
\alpha^{(n)}_g(t)\,.
\ee
Given that 
we work to NLL accuracy, 
we will only need the gluon Regge trajectory to first 
order in~$\as$, where in $d=4-2\eps$ dimensions
\be\label{GluonRegge1} 
\alpha^{(1)}_g(t) \,= \,\frac{\bn{0}(\eps)}{2\eps} 
\left(\frac{-t}{\mu^2}\right)^{-\eps} \,\,
\stackrel{\mu^2 \to -t}{=}\,\,\frac{\bn{0}(\eps)}{2\eps}\,.
\ee
Here, $\bn{0}(\eps)$ is a ubiquitous loop factor and 
the first of a class of bubble integrals, cf.\ \eqn{bubbleGeneral1}, 
to become important in section~\ref{soft}. For now, it suffices 
to know that
\be \label{eq:bzero}
\bn{0}(\eps) = e^{\eps\gamma_{\rm E}} 
\frac{\Gamma^2(1-\eps) 
\Gamma(1+\eps)}{\Gamma(1-2\eps)}
\,=\,1-\frac{\zeta_2}{2}\epsilon^2-\frac{7\zeta_3}{3}\epsilon^3+\ldots
.
\ee
In the following we will consider the leading contributions
to the signature-even amplitude to all orders, corresponding 
to the two-Reggeon exchange. These corrections ---  which 
we denote by ${\Mreduced}_{\rm NLL}^{(+)} $ --- were studied 
long ago~\cite{Kuraev:1977fs,Balitsky:1978ic} and can be 
expressed in terms the two-Reggeized-gluon wavefunction 
$\Omega(p,k)$ as follows: 
\be\label{ReducedAmpNLL}
{\Mreduced}_{\rm NLL}^{(+)} \left(\frac{s}{-t}\right) = -i\pi\int [\Dk] \, 
\frac{p^2}{k^2(p-k)^2}\, \Omega(p,k) \, \Tsu \, {\cal M}^{(\text{tree})}_{ij\to ij} \,,
\ee
where the integration measure is
\be\label{measure}
[\Dk] \equiv \frac{\pi}{B_0}
\, \left(\frac{\mu^2}{4\pi e^{-\gamma_E}}\right)^{\eps} 
\, \frac{\Dd^{2{-}2\eps}k}{(2\pi)^{2-2\eps}}\,.
\ee
with $\bn{0} \equiv \bn{0}(\eps)$ and the tree amplitude is
\beq
{\cal M}^{(\text{tree})}_{ij\to ij} =4\pi\alpha_s\,\frac{2s}{t} (T_i^b)_{a_1a_4}(T_j^b)_{a_2a_3}\delta_{\lambda_1\lambda_4}\delta_{\lambda_2\lambda_3}\,,
\eeq
where $\lambda_i$ for $i=1$ through $4$ are helicity indices.
The colour operator $\Tsu$ in \eqn{ReducedAmpNLL} acts on 
${\cal M}^{(\text{tree})}_{ij\to ij}$ and it is defined in terms of 
the usual basis of Casimirs corresponding to colour flow 
through the three channels~\cite{Dokshitzer:2005ig,DelDuca:2011ae}:
\beq\label{TtTsTu}
\Tsu \, \equiv \, \frac{\T_s^2-\T_u^2}{2}, 
\qquad  \rm{with} \qquad 
\left\{ \begin{array}{c}
\T_s =  \T_1+\T_2=-\T_3-\T_4,  \\ 
\T_u =  \T_1+\T_3=-\T_2-\T_4, \\
\T_t  =  \T_1+\T_4=-\T_2-\T_3,
\end{array} \right.\,
\eeq
where $\T_i$ represent the colour charge 
operator~\cite{Catani:1998bh} in the representation 
corresponding to parton $i$. The wavefunction 
$\Omega(p,k)$ has a perturbative expansion in 
the strong coupling, taking the form
\be \label{OmegaEven}
 \Omega(p,k)  = \sum_{\ell=1}^{\infty} \left(\frac{\alpha_s}{\pi}\right)^{\ell}
L^{\ell-1} \frac{B_0^{\ell}}{(\ell -1)!}\,\Wt{\ell-1}(p,k) \,,
\ee
where we set the renormalization 
scale equal to the momentum transfer, $\mu^2 = -t = p^2$.
The amplitude itself then has the corresponding expansion
\be\label{MhatEven}
\Mreduced_{\rm NLL}^{(+)}\left(\frac{s}{-t}\right) 
= \sum_{\ell=1}^\infty \left( \frac{\as}{\pi} \right)^\ell
\,L^{\ell -1}\, \Mreduced_{\rm NLL}^{(+,\ell)}\,.
\ee
We emphasise that while these corrections are the 
leading-logarithmic contributions to the even amplitude, 
we denote them by NLL to recall that the power of the 
logarithm $L$ is one less than the loop order. This can 
be contrasted with the single-Reggeized-gluon contribution 
to the odd amplitude $\MM_{\rm LL}^{(-)}\sim 
\,e^{\,\T_t^2\, \alpha_g(t) \, L} \, \Mtree$. 

In \eqn{MhatEven} $\Mreduced_{\rm NLL}^{(+,\ell)}$ 
contains $\ell$-loop diagrams and can be computed 
from the $(\ell-1)$-loop contribution to the wavefunction 
through integration
\be\label{ReducedAmpNLL2}
\Mreduced_{\rm NLL}^{(+,\ell)} = -i\pi\frac{(B_0)^\ell}{(\ell-1)!} \int [\Dk] \, 
\frac{p^2}{k^2(p-k)^2}\, \Wt{\ell-1}(p,k) \, \Tsu \, \Mtree\,.
\ee
In the normalisation used in \eqn{ReducedAmpNLL2}, 
the leading-order wavefunction is simply
\be
\label{0th-wavefunction-tilde}
\Wt{0}(p,k) = 1.
\ee
At loop level the wavefunction is then obtained 
iteratively by applying the BFKL Hamiltonian:
\bea 
\nn \Wt{\ell-1}(p,k) &=& (2C_A-\Tt)\,\int [\Dk'] \, f(p,k,k') 
\,\Wt{\ell-2}(p,k')+\tilde{J}(p,k)\, \Wt{\ell-2}(p,k)
\\ &\equiv& \hat{H} \,\Wt{\ell-2}(p,k)\label{Hdef}
\eea
where $f(p,k,k')$ is the BFKL evolution kernel 
\be\label{bfkl-kernel}
f(p,k,k') \equiv \frac{k^2}{k'^2(k - k')^2}
+\frac{(p-k)^2}{(p-k')^2(k - k')^2}
-\frac{p^2}{k'^2(p- k')^2}, 
\ee
and the function $\tilde{J}(p,k)$ is 
\be \label{Jp-def1}
\tilde{J}(p,k)  =  \frac{1}{2\eps}\left[
C_A\mratio{p}{k}^\eps+C_A\mratio{p}{(p-k)}^\eps-\Tt\right]\,.
\ee
\Eqn{Hdef} is the standard BFKL Hamiltonian 
(see eq.~(17) of the initial reference \cite{Kuraev:1977fs})
written using dimensional regularisation as an infrared 
regulator. $\tilde{J}(p,k)$ accounts for the Regge 
trajectories of the individual Reggeized gluons, minus 
the overall Regge trajectory with colour charge $\Tt$ 
which was subtracted in the exponent of the reduced 
amplitude (\ref{Mreduced}).

As discussed in refs.~\cite{Caron-Huot:2013fea,Caron-Huot:2017fxr} 
and briefly reviewed in appendix \ref{Review}, this equation 
and its higher-order generalisations can be understood by 
considering the expectation value of Wilson lines associated 
to the colour flow of the external partons \cite{Balitsky:1995ub}, 
which are described as ``target'' and ``projectile'' in the (high-energy) 
forward scattering configuration of figure~\ref{setup_fig}. 
The wavefunction then represents the transverse momenta 
in each of two Wilson lines and the BFKL equation is 
obtained as an appropriate limit of the more general 
Balitsky-JIMWLK evolution equation.

\begin{figure}[htb]
  \centering
 \includegraphics{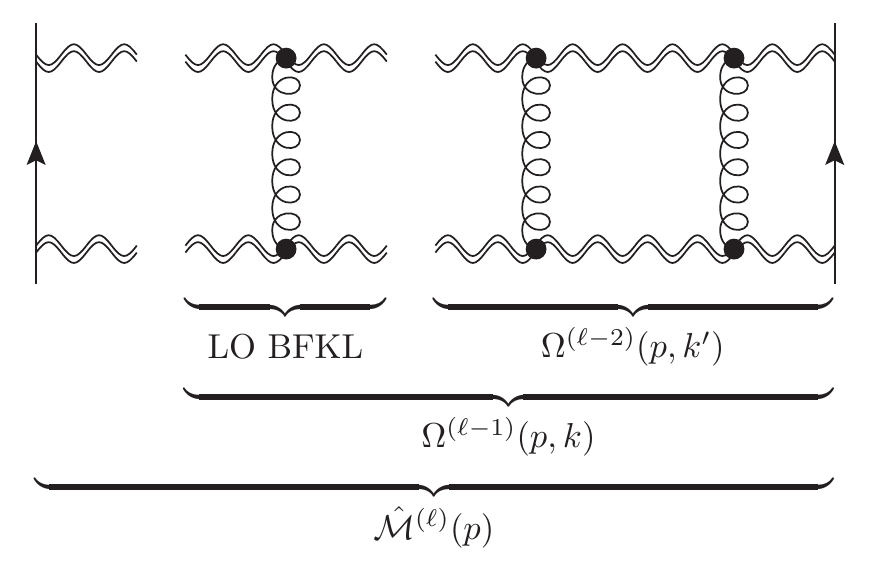}
  \caption{Graphical representation of the amplitude 
  at NLL accuracy, as obtained through BFKL evolution. 
  The addition of one rung corresponds to applying once 
  the leading-order BFKL evolution onto the projectile 
  wavefunction or impact factor at order $(\ell-2)$. 
  This gives the wavefunction at order $(\ell-1)$, 
  according to \eqn{Hdef1}. Closing the ladder and 
  integrating over the resulting loop momentum 
  gives the reduced amplitude, according to 
  \eqn{ReducedAmpNLL2}.} 
  \label{fig-ladder1}
\end{figure}

A graphical representation of 
\eqn{ReducedAmpNLL2} is 
provided in figure \ref{fig-ladder1}. As a 
result of BFKL evolution, the amplitude at NLL 
accuracy can be represented as a ladder.
At order $\ell$ it is obtained by closing the 
ladder and integrating the wavefunction 
of order $(\ell-1)$ over the resulting loop 
momentum, according to \eqn{ReducedAmpNLL2}. 
The wavefunction $\Wt{\ell-1}(p,k)$, in turn, is 
obtained by applying once the leading-order 
BFKL evolution kernel to the wavefunction of 
order $(\ell-2)$. Graphically, this operation 
corresponds to adding one rung to the ladder. 

\subsection{Iterative solution for the wavefunction and amplitude}
\label{multiloop-exact}

\Eqn{ReducedAmpNLL2} shows that the $\ell$-th order amplitude 
 is obtained in terms of iterated
integrals, which arise upon evaluating the 
wavefunction $\Wt{\ell-1}(p,k)$ to order $(\ell-1)$. 
It is straightforward to compute the first few orders, 
which gives us an opportunity to revisit the findings 
of ref.~\cite{Caron-Huot:2013fea}.
We will be able to explain why a new colour structure 
emerges for the first time at four loops, and explore 
the general structure of the relevant iterated integrals.

A useful fact is that the evolution admits one 
well-known solution in the case where the 
exchanged state is colour-adjoint and 
$\Omega(p,k)$ is constant (independent 
of $k$) \cite{Kuraev:1977fs,Balitsky:1978ic}, 
which gives a positive-signature state with 
the same leading-order trajectory as the 
Reggeized gluon. This enables one to 
rewrite the Hamiltonian (\ref{Hdef}) as a part 
which vanishes when $\Omega(p,k)$ is 
constant, plus a part proportional to $(C_A-\Tt)$:
\be \label{Hdef1}
\Wt{\ell-1}(p,k) =
\hat{H} \,\Wt{\ell-2}(p,k),
\qquad
 \hat{H} = (2C_A-\Tt) \,\hat H_{\rm i} + (C_A-\Tt) \, \hat H_{\rm m}
\ee
where, explicitly,
\bea \label{Him}\nn
 \hat H_{\rm i}\,\Psi(p,k) &=&\int [\Dk'] \, f(p,k,k') \Big[\Psi(p,k')-\Psi(p,k)\Big], \\
 \hat H_{\rm m}\,\Psi(p,k) &=& J(p,k)\, \Psi(p,k),
\eea
where the function $J(p,k)$ is defined by
\begin{align} \label{Jp-def2} \nn
J(p,k) &= \frac{1}{2\eps} + \int [\Dk'] \, f(p,k,k') \\
&= \frac{1}{2\eps} \left[2- \mratio{p}{k}^{\eps}-\mratio{p}{(p-k)}^{\eps}\right].
\end{align}

\begin{figure}[htb]
  \centering
  \includegraphics{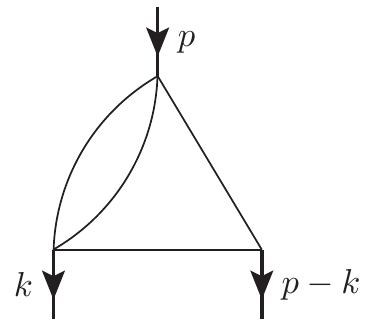}
  \caption{Three-mass triangle integral with 
  massless propagators appearing in the calculation of 
  the wavefunction at two loops. This type of integrals 
  contribute to the amplitude only starting at four loops, 
  due to the symmetry of the problem, as discussed in the 
  main text. The bubble integral on one of the two edges 
  of the triangles clarifies the origin of the propagator which 
  is raised to power $\eps$ in \eqn{triang1}.}
  \label{fig-triang1}
\end{figure}

The first interesting feature to note is that the $\hat H_{\rm i}$ operator
in \eqn{Hdef1} vanishes when acting on $\Wt{0}(p,k) = 1$. 
Therefore the wavefunction to one-loop involves a single 
colour structure:
\be
 \Wt{1}(p,k) = (C_A -\Tt) \, J(p,k)\,. \label{WavefunctionOneLoop}
\ee
The second colour structure appears 
for the first time at the second order:
\be \label{WavefunctionTwoLoops}
\Wt{2}(p,k) = (C_A-\Tt)^2 \left(J(p,k)\right)^2
+(2C_A-\Tt)(C_A-\Tt) \int [\Dk'] \, f(p,k,k')\Big[J(p,k')-J(p,k)\Big]\,.
\ee
Inserting the explicit form of $J(p,k)$ from 
\eqn{Jp-def2} into \eqn{WavefunctionTwoLoops}, 
one finds that it involves bubble integrals, as well 
as three-mass triangle integrals with massless 
propagators, such as 
\be \label{triang1}
\int[\Dk'] \,\frac{(p-k)^2}{(p-k')^2(k - k')^2} \left(\frac{p^2}{k'^2}\right)^{\eps},
\ee 
which is represented in figure~\ref{fig-triang1}. 
The wavefunction at higher orders can be expressed 
formally by introducing a class of functions
\bea \label{J_im_general}\nn
\Omega_{\rm i a_1 \ldots a_n}(p,k) &\equiv& 
\int [\Dk'] \, f(p,k,k') \Big[\Omega_{\rm a_1\ldots a_n}(p,k')-\Omega_{\rm a_1\ldots a_n}(p,k)\Big]\,, \\ 
\Omega_{\rm m  a_1 \ldots a_n}(p,k)  &\equiv& J(p,k) \, \Omega_{\rm a_1 \ldots a_n}(p,k)\,,
\eea
where $\Omega_{\varnothing}(p,k)\equiv 1$, and 
each of the indices ${\rm a_j}$ can take the value
``i'' or ``m'', which stand for integration and multiplication, 
respectively, according to the action of the two Hamiltonian 
operators in \eqn{Him}.
In this notation, the one- and two-loop wavefunctions read, respectively,
\bea \label{WavefunctionTwoLoops-b} \nn
 \Wt{1}(p,k) &=& (C_A -\Tt) \,\Omega_{\rm m}\,,\\
 \Wt{2}(p,k) &=& (C_A-\Tt)^2 \,\Omega_{\rm m m}
 +(2C_A-\Tt)(C_A-\Tt) \,\Omega_{\rm i m}\,,
\eea
and it is also easy to write the wavefunctions at higher loops, for example:
\bea \nn
 \Wt{3}(p,k) &=& (C_A-\Tt)^3 \,\Omega_{\rm mmm} 
 + (2C_A-\Tt)(C_A-\Tt)^2 \big(\Omega_{\rm i m m}+\Omega_{\rm m i m}\big)
\\ &+&  (2C_A-\Tt)^2(C_A-\Tt) \, \Omega_{\rm i i m}\,.
\eea

The wavefunctions written thus far are sufficient 
to evaluate the reduced amplitude up to four loops.  
At one and two loops, inserting respectively 
$\Wt{0}(p,k) = 1$ and \eqn{WavefunctionOneLoop} 
into \eqn{ReducedAmpNLL2} and performing 
bubble integrals one gets immediately
\bea \label{ReducedAmpNLL2-one-two-loop}
\Mreduced_{\rm NLL}^{(+,1)} &=& - i\pi \,
\frac{\bzero}{2\eps} \, \Tsu \, \Mtree , \\
\Mreduced_{\rm NLL}^{(+,2)} &=& i\pi \,
\frac{(\bzero)^2}{2}\left[ \frac{1}{(2\eps)^2} 
+ \frac{9\zeta_3}{2}\eps + \frac{27\zeta_4}{4}\eps^2 
+ \frac{63\zeta_5}{2}\eps^3+ \ord(\eps^4)\right]  
\,  (C_A -\Tt)\, \Tsu \, \Mtree.\nn
\eea
We notice that the amplitude depends
solely on the colour structure $(C_A -\Tt)$, 
and this in turn is a consequence of the 
fact that the wavefunctions $\Wt{0}$
and $\Wt{1}$ have only one colour component.
Based on this consideration alone,
one would expect the second colour structure, $(2C_A -\Tt)$, 
to contribute to the amplitude starting at three 
loops, given that it appears in $\Wt{2}(p,k)$ of 
eq.~(\ref{WavefunctionTwoLoops-b}). However, 
this contribution of  $\Wt{2}(p,k)$ to the amplitude 
$\Mreduced_{\rm NLL}^{(+,3)}$ cancels by 
symmetry: 
\bea \label{JiJmSym}\nn
\int [\Dk] \, \frac{p^2}{k^2(p-k)^2} \, \Omega_{\rm i m}(p,k)
&=&
\int [\Dk] \, [\Dk'] \, \frac{p^2}{k^2(p-k)^2}\,f(p,k,k')\Big[  J(p,k') - J(p,k) \Big] \\
&&\hspace{-30mm}=\,
\int [\Dk] \, [\Dk']  \bigg\{ \frac{p^2}{k'^2(p-k')^2}\, f(p,k',k) J(p,k') - (k\leftrightarrow k') \bigg\}= 0,
\eea
where in the last line we used the 
property 
\be \label{f-prop}
\frac{p^2}{k'^2(p-k')^2}\, f(p,k',k) = \frac{p^2}{k^2(p-k)^2}\, f(p,k,k'),
\ee
which makes 
evident that \eqn{JiJmSym} vanishes by 
antisymmetry with respect to $k \leftrightarrow k'$.
Because of this, the amplitude at three loops 
has again a single colour component, proportional to $(C_A -\Tt)^2$: 
\be \label{ReducedAmpNLL2-three-loops}
\Mreduced_{\rm NLL}^{(+,3)} = i\pi \,
\frac{(\bzero)^3}{3!} \left[ \frac{1}{(2\eps)^3} - \frac{11\zeta_3}{4}
- \frac{33\zeta_4}{8}\eps - \frac{357\zeta_5}{4}\eps^2 
+ \ord(\eps^3) \right]\,  (C_A -\Tt)^2 \, \Tsu \, \Mtree.  
\ee
This symmetry relation
generalises to higher orders, i.e.\ one has 
\be \label{JiJmSymAllOrders}
\int [\Dk] \, \frac{p^2}{k^2(p-k)^2}\Omega_{\rm i a_1 \ldots  a_n}(p,k) = 0,
\ee
for any ${\rm a_1 \ldots a_n}$. 
While this symmetry ensures that there is only 
one colour structure at three loops, this is no 
longer the case starting at four loops. There, 
one obtains \cite{Caron-Huot:2013fea}
\bea \label{ReducedAmpNLL2-four-loops} \nn
\Mreduced_{\rm NLL}^{(+,4)} &=&  - i\pi 
\frac{(\bzero)^4}{3!} \int [\Dk] 
\frac{p^2}{k^2(p-k)^2} \bigg\{ 
(C_A -\Tt)^3 \,\Omega_{\rm mmm}(p,k) \\ \nn&&\hspace{4.0cm}
+ \,(2C_A-\Tt)(C_A-\Tt)^2 \,\Omega_{\rm mim}(p,k)\bigg\} \, \Tsu \, \Mtree \\ 
&=&i\pi \,
\frac{(\bzero)^4}{4!} \bigg\{ 
(C_A -\Tt)^3 \bigg( \frac{1}{(2\eps)^4}  + \frac{175\zeta_5}{2}\eps  + {\cal O}(\eps^2)\bigg) \\ \nn
&&\hspace{0.9cm}+\, C_A (C_A -\Tt)^2
\bigg( -\frac{\zeta_3}{8\eps} - \frac{3}{16}\zeta_4 - \frac{167\zeta_5}{8}\eps
+ \ord(\eps^2) \bigg) \bigg\} \, \Tsu \, \Mtree. 
\eea
One sees that the integrated result involves two 
colour structures, and in the final expression in 
\eqn{ReducedAmpNLL2-four-loops} we rearranged 
them so as to single out a factor of $C_A$. 
In section \ref{IRfact} below we will see that 
in this form it is easy to compare the amplitude with 
the structure of infrared divergences. Specifically, 
we will see that corrections involving the colour 
structure $(C_A -\Tt)^{\ell-1}$ at $\ell$ loop order 
emerge directly from the simplest ``dipole'' formula 
of the soft anomalous dimension, while other colour 
structures, namely $C_A^j (C_A -\Tt)^{\ell-j-1}$ with 
$j\geq1$, identify deviations from the dipole formula, 
as was first observed in ref.~\cite{Caron-Huot:2013fea}
for $\ell = 4$. 
\begin{figure}
  \centering
 \includegraphics{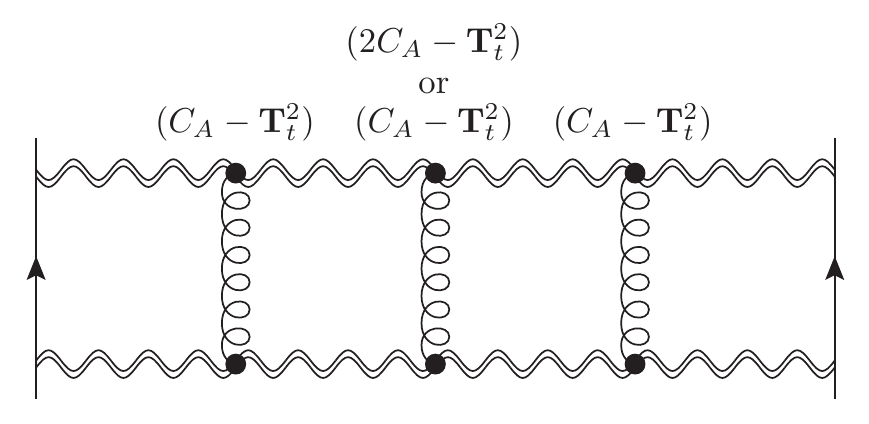}
  \caption{Graphical representation of the BFKL ladder
  at four loops. The fact that $\Wt{1}(p,k) \sim (\CA-\Tt)$ in 
  conjunction with the target-projectile symmetry imply that 
  the first rungs on either side can only give rise to contributions 
  proportional to $(\CA-\Tt)$. As a consequence, distinct
  colour structures can appear for the first time at four loops.}
  \label{fig-ladder4L}
\end{figure}

Inspecting the diagrammatic representation of 
BFKL evolution in figure~\ref{fig-ladder1}, one can  
interpret the delayed appearance of a new colour 
structure to four loops, as a consequence of the 
target-projectile symmetry. Recall that for the first 
rung of the ladder, only the second term 
$\hat{H}_{\rm m}$ in \eqn{Hdef1} contributes, 
so the wavefunction has a single colour structure 
$(C_A - \Tt)$. Considering more rungs, using 
target-projectile symmetry one can deduce that 
the same is true for the first rung on the opposite 
side of the ladder. As a consequence, despite the 
fact that $\Wt{2}(p,k)$ contains two structures 
(see \eqn{WavefunctionTwoLoops-b}), the effect 
of the second one, $(2C_A - \Tt)$, cannot appear 
in the three-loop amplitude, where each of the 
two rungs contribute a factor of $(C_A - \Tt)$. 
As shown in figure~\ref{fig-ladder4L}, distinct colour 
structures can only appear in the 
amplitude starting at four loops, where the middle rung 
--- and only that rung --- gives rise to both colour factors.

\section{The soft approximation}
\label{soft}

While it would be possible to calculate
the wavefunction and amplitude to higher loop orders, 
in this paper we focus on the infrared divergent part 
of the latter. We strive to compare its singularities with 
the predictions made by the infrared factorisation 
theorem and, consequently, deduce higher-order 
corrections to the high-energy soft anomalous 
dimension. With this goal in mind, we highlight 
at this point another important property of 
$\Omega(p,k)$, which can be verified when 
inspecting \eqn{Hdef1} more carefully (see below): 
the wavefunction $\Omega^{(\ell -1)}(p,k)$ is finite 
for $\eps \to 0$ to all orders in perturbation theory!
This is a non-trivial statement, which becomes 
evident only after the evolution evolution equation 
is brought from the form in \eqn{Hdef} to \eqn{Hdef1}. 
A practical implication is that all divergences in the 
amplitude must originate in the final integration, 
namely going from the wavefunction to the amplitude 
as in \eqn{ReducedAmpNLL}. Inspecting the latter 
equation, we see that divergences arise only in the 
$k \to 0$ and $k \to p$ limits (and ultraviolet power 
counting in \eqn{Him} using \eqref{bfkl-kernel} 
excludes divergences from $k'\gg p,k$). Due to 
the symmetry of the integrand, all divergences of 
the amplitude can therefore be obtained by 
evaluating it in one of these two limits, and 
multiplying the result by two. 

Let us now examine more carefully the evolution of 
the wavefunction according to eqs.~(\ref{Hdef1}) 
and (\ref{Him}), verify that the wavefunction is indeed 
finite, and derive a simplified version of the evolution, 
valid in the small-$k$ or \emph{soft} approximation: $k \ll p$.
The loop integral in \eqn{Him} can in  principle receive 
contributions from two regions; $k \ll k' \sim p$ and 
$k \sim k' \ll p$. Inspecting the form of $f(p,k,k')$ in 
the two regions, it is easy to check that only the 
second region contributes: 
\bea \nn
f(p,k,k')|_{k \ll k' \sim p} &\longrightarrow  & 
0+\frac{p^2}{(p-k')^2 k'^2}
-\frac{p^2}{k'^2(p- k')^2}\,= \, 0, \\
f(p,k,k')|_{k \sim k' \ll p} &\longrightarrow& 
\frac{k^2}{k'^2(k - k')^2}
+\frac{1}{(k - k')^2} -\frac{1}{k'^2} = 
\frac{2(k\cdot k')}{k'^2(k - k')^2}.
\eea
This means that the soft approximation 
closes under evolution! In the following, 
we will identify the  region $k \sim k' \ll p$ 
as \emph{soft} and add a subscript $s$ 
to quantities calculated in this limit. 
From $J(p,k)$ in \eqn{Jp-def2} one 
gets
\be\label{JpSoft}
J_s(p,k) = \frac{1}{2\eps} \left[1-\mratio{p}{k}^{\eps}\,\right], 
\ee
and the evolution in \eqn{Hdef1} becomes 
\begin{samepage}
\bea \nn
 \Omega^{(\ell-1)}_s(p,k)&=&\hat H_s \, \Omega^{(\ell-2)}_s(p,k)\,, \\ \nn
\hat H_s \Psi(p,k) &=& (2C_A-\Tt)\int[\Dk']
\frac{2(k\cdot k')}{k'^2(k - k')^2} \, \Big[ \Psi(p,k')-\Psi(p,k)\Big] 
\\&& 
+\, (C_A-\Tt)\, J_s(p,k) \, \Psi(p,k)\,, \label{softH}
\eea
 where $[\Dk']$ is the previously defined 
integration measure \eqref{measure}.
\Eqn{softH} confirms that it is consistent 
to truncate the Regge evolution to the soft 
approximation: using the power counting 
$\Psi(p,k)\sim 1$, we see that the $k'$ integral 
is saturated by the soft region $k'\sim k$, with 
no sensitivity to larger scales.
\end{samepage}

Inserting the wavefunction $\Omega_s^{(\ell-1)}(p,k)$ into 
\eqn{ReducedAmpNLL2}, we get the amplitude in the soft 
limit at the $\ell$-th order. In this approximation the last integral 
becomes divergent and needs an ultraviolet cutoff, which 
we fix by requiring $k^2<p^2$, based on dimensional 
analysis and consistency with the soft limit (any cutoff 
would be consistent, and would not affect the infrared 
singularities). The integration measure for the last integral 
therefore reads 
\be\label{int-measure-soft}
\int [Dk ]_s = \frac{(p^2)^{\eps} \, 
e^{\eps \gamma_E}}{2 \Gamma(1-\eps) B_0} 
\int_0^{p^2} \dk^2 (k^2)^{-\eps},
\ee
where we multiplied by a factor of two, in order to take
into account the fact that there is an identical contribution 
from the region where the Reggeized gluon carrying momentum $(p-k)$ is soft. 
Inserting this result into \eqn{ReducedAmpNLL2}, we get 
$\Mreduced_{\rm NLL}^{(+,\ell)}$ in the soft approximation: 
\be\label{MellReggeSoft}
\Mreduced_{\rm NLL}^{(+,\ell)} =  - \frac{i\pi (B_0)^{\ell-1}}{(\ell-1)!} \,
\frac{e^{\eps \gamma_E}}{\Gamma(1-\eps)} 
\int_0^{p^2} \frac{\dk^2}{2k^2} \, \mratio{p}{k}^{\eps}  
\,\Wts{\ell-1}(p,k) \, \Tsu \, \Mtree +{\cal O}(\eps^{0}).
\ee
We stress that this approximation gives correct results 
only as far as infrared singularities are concerned.
All poles in $\eps$ are exact, since the integrand is 
finite and divergences arise only from the $k \to 0$ 
limit of integration. The reduced amplitude in 
\eqn{MellReggeSoft} ceases to be correct at 
finite ${\cal O}(\eps^{0})$ order, as indicated.

The most significant advantage of the 
soft approximation is that the evolution 
equation greatly simplifies, and this allows 
us to obtain closed-form expressions for the 
wavefunction $\Wts{\ell-1}(p,k)$ and the 
amplitude $\Mreduced_{\rm NLL}^{(+,\ell)}|_{s}$, 
as we are going to detail in the following.

\subsection{The wavefunction at NLL to all orders}
\label{WsoftAllOrders}

In analogy to the exercise done in section \ref{multiloop-exact}, 
we start by calculating explicitly the wavefunction at
the first few orders in perturbation theory, this time in 
the soft approximation. The initial condition is still given 
by \eqn{0th-wavefunction-tilde}, and the evolution 
obeys \eqn{softH}. This equation has a much 
simpler structure compared to the original one, 
\eqn{Him}, because the soft approximation turns 
a two-scale problem into a one-scale problem. It is easy 
to check that the wavefunction reduces to a polynomial 
in $\xi = \left(p^2/k^2\right)^{\eps}$, which implies that the 
integrals involved in \eqn{softH} are
simple bubble integrals of the type 
\be\label{bubbleGeneral1}
\int [\Dk'] \, \frac{2(k\cdot k')}{k'^2(k - k')^2} \, 
\mratio{p}{k'}^{n \eps} =
-\frac{1}{2\eps}\frac{\bn{n}(\eps)}{\bn{0}(\eps)}\mratio{p}{k}^{(n+1)\eps}\,,
\ee
where the integration measure is given in eq.~(\ref{measure}).
This defines a class of one-loop functions mentioned 
above \eqn{eq:bzero}, namely
\be \label{bubbleGeneral2}
\bn{n}(\eps) = e^{\eps\gamma_{\rm E}}  \frac{\Gamma(1-\eps)}{\Gamma(1+n\eps)}
\frac{\Gamma(1+\eps + n\eps) \Gamma(1-\eps - n\eps)}{\Gamma(1-2 \eps - n\eps)}\,.
\ee
Using this we can write the action of the soft Hamiltonian 
(\ref{softH}) on any monomial ($m\geq 0$):
\bea
 \label{softHpower}
 \hat H_s \,\xi^m &=& \frac{\xi^m}{2\eps} \left((1-\xi)(C_A-\Tt) 
 + \xi \hbn{m}(\eps)(2C_A-\Tt)\right)\,  \\
&=& \frac{(C_A-\Tt)}{2\eps} \left(\xi^m - \xi^{m+1}
\left[ 1- \hbn{m}(\eps)\frac{2C_A-\Tt}{C_A-\Tt}\right]\right)\,,\nn
\eea
where we have introduced the loop functions
\be\label{bubblehat} 
\hat B_n(\eps) = 1- \frac{B_n(\eps)}{\bzero(\eps)}
\,=\,
2 n (2 + n) \zeta_3 \epsilon^3 
+ 3 n (2 + n) \zeta_4 \epsilon^4 
+\ldots.
\ee
Given that $\hat B_m(\eps)={\cal O}(\epsilon^3)$, the 
first line in \eqn{softHpower} makes manifest the fact 
that $\hat H_s \,\xi^m$ is finite for $\epsilon\to 0$, in 
line with our earlier assertion about the finiteness of 
the wavefunction. The second line will be useful in 
what follows for determining the all-order structure 
of the wavefunction.

Applying  \eqn{bubbleGeneral1} repeatedly up to 
three loops (which is sufficient to determine the 
amplitude at four loops) we find 
\bea 
\label{soft_wavefunction_first_orders}
\Wts{0}(\xi) &=& 1, \\ \nn 
\Wts{1}(\xi) &=& \frac{(C_A-\Tt)}{2\eps} 
\Big(1-\xi\Big), \\ \nn 
\Wts{2}(\xi) &=& \frac{(C_A-\Tt)^2}{(2\eps)^2} 
\bigg\{1-2\xi+ \xi^2\left[ 1- \hbn{1}(\eps)\frac{2C_A-\Tt}{C_A-\Tt}\right]\bigg\}\,,\\ \nn 
\Wts{3}(\xi) &=& \frac{(C_A-\Tt)^3}{(2\eps)^3} 
\bigg\{1-3\xi + 3\xi^2\left[ 1- \hbn{1}(\eps)\frac{2C_A-\Tt}{C_A-\Tt}\right] \\ \nn
&& \hspace{25mm}-\,\xi^3\left[ 1- \hbn{1}(\eps)\frac{2C_A-\Tt}{C_A-\Tt}\right]
\left[ 1- \hbn{2}(\eps)\frac{2C_A-\Tt}{C_A-\Tt}\right]\bigg\}.
\eea
The evaluation of a few additional orders allows us to
obtain an ansatz for the $(\ell-1)$-th order wavefunction:
\be \label{Well-1-ansatz}
\Wts{\ell-1}(p,k) =
\frac{(\CA-\Tt)^{\ell-1}}{(2\eps)^{\ell-1}} \sum_{n=0}^{\ell-1} 
(-1)^n \binom{\ell-1}{n} \mratio{p}{k}^{n\eps} 
\prod_{m=0}^{n-1} 
\left\{1 - \hbn{m}(\eps) \frac{2\CA-\Tt}{\CA-\Tt}\right\}\,.
\ee
The validity of this all-order formula can be proved directly 
using the action of the Hamiltonian in the second line of 
\eqn{softHpower} by noticing first that, independently of 
the loop order, the term $\xi^n$ can only be generated by 
acting $n$ times with the second term of \eqn{softHpower}, 
each of which raises the power of $\xi$ by one. Hence 
$\xi^n$ will always be accompanied by the product
$(-1)^n\prod_{m=0}^{n-1} 
\left\{1 - \hbn{m}(\eps) \frac{2\CA-\Tt}{\CA-\Tt}\right\}$.
Furthermore, the combinatorial factor $\binom{\ell-1}{n}$ 
associated with $\xi^n$ simply counts the number of different 
ways of acting $(\ell-1)$ times with the Hamiltonian, out of 
which $n$ times with the second term and $\ell-1-n$ times 
with the first.

\subsection{The all-order structure of two-parton scattering amplitudes at NLL}
\label{M-all-orders-soft}

The main result of the previous section 
is that, in the soft 
approximation, the wavefunction reduces to 
a polynomial in $\left(p^2/k^2\right)^{\eps}$, 
given by~\eqn{Well-1-ansatz}. As a consequence, 
the calculation of the amplitude 
\eqref{MellReggeSoft} becomes 
straightforward, because it involves 
only integrals of the type 
\be \label{lastint}
\int_0^{p^2} \frac{\dk^2}{k^2} \,
\mratio{p}{k}^{n\, \eps} = -\frac{1}{n \, \eps},
\ee
which allows us to obtain 
\begin{multline} \label{MellReggeSoft-ResB}
\left.\Mreduced_{\rm NLL}^{(+,\ell)}\right|_{s} = i \pi \, 
\frac{1}{(2\eps)^{\ell}} \, 
\frac{\bzero^{\ell}(\eps)}{\ell!} \, (1-\hbn{-1})
\,  (C_A -\Tt)^{\ell-1} \sum_{n=1}^{\ell} 
(-1)^{n+1} \, \binom{\ell}{n} \\
\times \prod_{m=0}^{n-2}\bigg[ 1 - \hbn{m}(\eps) \frac{2 C_A -\Tt}{C_A -\Tt} \bigg]
 \, \Tsu \, \Mtree +{\cal O}(\eps^{0}),
\end{multline}
where the factor $(1-\hbn{-1})$ follows from 
rewriting the factor $e^{\eps \gamma_E}/\Gamma(1-\eps)=\bn{-1}(\epsilon)$:
\be
(\bzero)^{\ell-1} \, \frac{e^{\eps \gamma_E}}{\Gamma(1-\eps)} =
(\bzero)^{\ell} \frac{\bn{-1}(\eps)}{\bn{0}(\eps)} = 
(\bzero)^{\ell} (1-\hbn{-1}).
\ee
\Eqn{MellReggeSoft-ResB}
looks rather involved but one must keep in 
mind that, upon expansion in $\eps$, it contains 
many finite terms which do not represent the actual 
amplitude since we are working in the soft 
approximation. Given the overall factor of 
$1/(2\eps)^{\ell}$ in \eqn{MellReggeSoft-ResB}, 
all the singularities are obtained by retaining only 
contributions up to $\eps^{\ell-1}$ in the subsequent 
factors. When this is taken into account a great 
simplification arises: indeed, as shown in 
appendix \ref{AppB}, it is possible to 
prove that 
\eqn{MellReggeSoft-ResB} is equivalent to 
\begin{multline} \label{MellReggeSoft-All-B}
\left.\Mreduced_{\rm NLL}^{(+,\ell)}\right|_{s}=  i \pi \,  
\frac{1}{(2\eps)^{\ell}} \, 
\frac{\bzero^{\ell}(\eps)}{\ell!} \, (1-\hbn{-1})
\left( 1- \hbn{-1}(\eps) \frac{2C_A-\Tt}{C_A -\Tt} \right)^{-1} \\
\times (C_A -\Tt)^{\ell-1} \, \Tsu \, \Mtree +{\cal O}(\eps^{0}).
\end{multline}
It is remarkable that the complicated sum of 
products of bubble integrals weighed by a 
binomial factor collapses to a single factor 
which depends only on one bubble integral, 
namely $\hbn{-1}(\eps)$. The main ingredient 
of the proof is the fact that the wavefunction 
itself is finite.

\Eqn{MellReggeSoft-All-B} constitutes the main
result of this section: by iterating the BFKL
equation (which was not diagonalised before 
in $d=4-2\eps$ dimensions) we obtained the 
singular part of  the even amplitude at NLL 
accuracy, to all orders in the strong coupling 
constant. Anticipating comparison with the 
structure of infrared divergences dictated by 
the soft anomalous dimension, it proves useful 
to rearrange \eqn{MellReggeSoft-All-B} in such 
a way to single out the colour structures $C_A$ 
and $(C_A -\Tt)$. Indeed, as discussed at the 
end of section \ref{multiloop-exact}, we know that 
the dipole formula of infrared divergencies fixes 
the singularities of the even amplitude in the 
high-energy limit to be proportional to the 
colour structure $(C_A -\Tt)^{\ell-1}\Tsu$ 
at $\ell$ loops. From \eqn{MellReggeSoft-All-B} 
we obtain 
\be\label{MellReggeSoft-All-C}
\left.\Mreduced_{\rm NLL}^{(+,\ell)}\right|_{s}=  i \pi \,  
\frac{1}{(2\eps)^{\ell}} \, 
\frac{\bzero^{\ell}(\eps)}{\ell!} \, 
\left( 1 - R(\eps) \frac{C_A}{C_A -\Tt} \right)^{-1} 
(C_A -\Tt)^{\ell-1} \, \Tsu \, \Mtree +{\cal O}(\eps^{0}),
\ee
where we have introduced the function 
\bea \nn
R(\eps) \equiv \frac{\bzero(\eps)}{\bn{-1}(\eps)} -1 &=& 
\frac{\Gamma^{3}(1-\eps)\Gamma(1+\eps)}{\Gamma(1-2\eps)} -1  \\
\label{Rdef}
&=& -2\zeta_3 \, \eps^3 -3\zeta_4 \, \eps^4 -6\zeta_5 \eps^5
-\left(10 \zeta_6-2\zeta^2_3 \right) \eps^6 + {\cal O}(\eps^7).
\eea
Furthermore, by resumming \eqn{MellReggeSoft-All-C} 
according to \eqn{MhatEven} we get the all-order amplitude: 
\begin{multline} \label{eq:mredregge}
  \left.\Mreduced_\NLL^{(+)}\right|_s = \frac{i\pi}{L(\CA-\Tt)} 
 \left( 1 - R(\eps) \frac{C_A}{C_A -\Tt} \right)^{-1} \\
  \times \left[ \exp\left\{\frac{\bzero(\eps)}{2\eps} 
  \frac{\as}{\pi} L (\CA-\Tt)\right\} - 1 \right] \Tsu \, \Mtree + \ord(\eps^0).
\end{multline}
This result will be used in the next section to 
extract the soft anomalous dimension. 

Before 
addressing this topic, however, it proves useful 
to explore in more detail the implications of 
\eqn{MellReggeSoft-All-C} by writing explicitly 
a few orders in perturbation theory. 
Up to three 
loops \eqn{MellReggeSoft-All-C} reduces to
\be\label{MellReggeSoft123} 
\left.\Mreduced_{\rm NLL}^{(+,\ell=1,2,3)}\right|_{s} = i \pi \, 
\frac{\bzero^{\ell}(\eps)}{\ell! \, (2\eps)^{\ell}} 
\,  (C_A -\Tt)^{\ell-1} \,  \Tsu \, 
\Mtree +{\cal O}(\eps^{0}), 
\ee
i.e.\ only one colour structure 
contributes to the amplitude up to three
loops, and the singularities are correctly 
reproduced by the dipole formula of infrared 
divergences. Starting at four loops, and 
for the subsequent three orders, one 
gets an additional contribution proportional 
to a new colour structure: 
\be \label{MellReggeSoft456} 
\left.\Mreduced_{\rm NLL}^{(+,\ell=4,5,6)}\right|_{s} = i \pi \,
\frac{\bzero^{\ell}(\eps)}{\ell! \,(2\eps)^{\ell}} \bigg\{ (C_A -\Tt)^{\ell-1} 
+ R(\eps) \, C_A (C_A -\Tt)^{\ell-2} 
\bigg\}\, \Tsu \, \Mtree + {\cal O}(\eps^{0}),
\ee
which matches with the infrared-divergent part of 
the result reported earlier in \eqn{ReducedAmpNLL2-four-loops}.
It can be easily verified (see the next section) that the 
infrared divergences associated with the first colour 
structure are predicted by the dipole formula, while 
the ones associated with the second are not. Next, 
starting at seven loops, and for the subsequent 
three orders, yet another colour structure arises: 
\bea \label{MellReggeSoft789}
\left.\Mreduced_{\rm NLL}^{(+,\ell = 7,8,9)}\right|_{s} &=& i \pi \,
\frac{\bzero^{\ell}(\eps)}{\ell! \,(2\eps)^{\ell}} \bigg\{ 
(C_A -\Tt)^{\ell-1} + R(\eps)\, C_A (C_A -\Tt)^{\ell-2} \\  
&&\hspace{2.0cm} + \, R^{2}(\eps) \, C_A^2 (C_A -\Tt)^{\ell-3}\bigg\}\, 
\Tsu \, \Mtree + {\cal O}(\eps^{0})\,. \nn
\eea
Expanding \eqn{MellReggeSoft-All-C}
for the next three orders in $\as$ we get 
\bea \label{MellReggeSoft101112}
\left.\Mreduced_{\rm NLL}^{(+,\ell = 10,11,12)}\right|_{s} &=& i \pi \,
\frac{\bzero^{\ell}(\eps)}{\ell! \,(2\eps)^{\ell}} \bigg\{ 
(C_A -\Tt)^{\ell-1} + R(\eps)\, C_A (C_A -\Tt)^{\ell-2}  \\ 
&&\hspace{-2.5cm} +\, R^{2}(\eps)\,  C_A^2 (C_A -\Tt)^{\ell-3} 
+ R^{3}(\eps)\, C_A^3 (C_A -\Tt)^{\ell-4} \bigg\}\, 
\Tsu \, \Mtree + {\cal O}(\eps^{0})\,. \nn
\eea
It is now easy to understand the 
pattern singularities implied by \eqn{MellReggeSoft-All-C}: 
at each order
the first colour structure, proportional to $(C_A -\Tt)^{\ell-1}$, 
describes the singularities predicted by the dipole formula. 
Additional colour structures are generated by the expansion 
of the geometric series $1/\left( 1 - R(\eps) \frac{C_A}{C_A -\Tt} \right)$ 
in \eqn{MellReggeSoft-All-C}, such that 
every three loops a new colour structure 
arises with an increasing power of $C_A$, 
replacing one of the factors of $(C_A-\Tt)$. All 
these new structures introduce infrared divergences, 
which are not accounted for by the dipole formula. 

Now that we understand the result implied by the
BFKL evolution equation, we are in the position to 
investigate how the infrared divergences not 
accounted for by the dipole formula can be 
included in the soft anomalous dimension. This 
will be the subject of the following section.

\section{The soft anomalous dimension in the high-energy limit to all orders}\label{IRfact}

It is well known that infrared divergences 
in gauge-theory scattering amplitudes are 
multiplicatively ``renormalizable'': finite 
hard-scattering amplitudes may be obtained 
by multiplying the original infrared-divergent
amplitude by a renormalization factor 
$\Z(\{p_i\},\mu,\as(\mu))$, which is matrix-valued 
in colour-flow space. This factor solves a 
renormalization group equation, and hence 
can be written as a path-ordered exponential 
of a soft anomalous dimension 
${\bf \Gamma}(\{p_i\},\mu,\as(\mu))$, 
integrated over the scale $\mu$. As such, the 
soft anomalous dimension constitutes a 
fundamental ingredient for the calculation 
of scattering processes at any given order 
in perturbation theory, and much effort has 
been devoted to its determination. It has 
been shown that the soft anomalous 
dimension has a simple dipole 
structure up to two loops~\cite{Aybat:2006mz}. 
Corrections involving three and four partons 
arise starting at three loops, and a series of 
analyses has been performed in order to 
constrain their structure at three loops and 
beyond \cite{Becher:2009cu,Gardi:2009qi,Becher:2009qa,Gardi:2009zv,Dixon:2009ur,Ahrens:2012qz};
the complete correction at three loops was 
calculated recently~\cite{Almelid:2015jia,Gardi:2016ttq}.

The general structure of the soft anomalous 
dimension is fixed by the factorisation properties 
of soft and collinear radiation, along with symmetry 
properties, such as rescaling invariance of soft 
corrections with respect to the momenta of the 
hard partons. The latter properties link dipole 
terms to the cusp anomalous dimension and 
dictate the structure of corrections to the soft 
anomalous dimension that correlate more than 
two partons \cite{Becher:2009cu,Gardi:2009qi,Becher:2009qa,Gardi:2009zv}. 
In particular, they imply that at three loops, 
non-dipole corrections can only depend on the kinematics via
rescaling-invariant cross ratios. The soft 
anomalous dimension can be further 
constrained by the behaviour of scattering 
amplitudes in special kinematic limits, 
such as the Regge limit~\cite{Bret:2011xm,DelDuca:2011ae,Caron-Huot:2017fxr} 
and collinear limits~\cite{Becher:2009cu,Dixon:2009ur}. 
Furthermore, it was recently shown~\cite{Almelid:2017qju} 
that the space of functions in terms of which the 
non-dipole correction is expressed (single-valued 
multiple polylogarithms) can, in fact, be deduced 
from general considerations. A bootstrap procedure 
was then set up, which remarkably completely fixes 
the functional form of the non-dipole correction at three loops
(up to an overall rational numerical factor) based on 
known information from the kinematic limits 
mentioned above, reproducing the result of 
the Feynman-diagram computation of 
ref.~\cite{Almelid:2015jia,Gardi:2016ttq}. 
The prospects of extending this bootstrap 
procedure to higher loops provides an 
additional motivation to determining the 
soft anomalous dimension in the 
high-energy limit.

As discussed above, ref.~\cite{Caron-Huot:2013fea} 
determined the next-to-leading high-energy logarithms 
(NLL) of $2 \to 2$ scattering amplitudes at four loops.
In this paper we have been able to extend this and 
computed the infrared singularities at NLL in the high-energy limit 
to all order in perturbation theory. We are therefore able to determine 
the soft anomalous dimension in this approximation to all orders.

We start this section by briefly reviewing the 
structure of the soft anomalous dimension in 
the high-energy limit, and then determine it to 
all orders by extracting the ${\cal O}(1/\epsilon)$ 
coefficient from the amplitude obtained in section 
\ref{M-all-orders-soft}, which we then analyze numerically in detail.
Finally we show that the 
singularity structure we deduced from the high-energy 
limit computation, consisting of poles of 
${\cal O}(1/\epsilon)$ through to ${\cal O}(1/\epsilon^\ell)$ 
at $\ell$ loops, is consistent with infrared factorisation, 
namely it is exactly reproduced by the expansion of 
the path-ordered exponential of the integral of the 
soft anomalous dimension.

\subsection{The infrared factorisation formula in the Regge limit} 
\label{sec:IR_fact}

The infrared divergences of scattering amplitudes 
can be factorised as
\beq \label{IRfacteq}
\MM \left(\{p_i\},\mu, \as (\mu) \right) \, = \, 
{\bf Z} \left(\{p_i\},\mu, \as (\mu) \right)
\Hhard \left(\{p_i\},\mu, \as (\mu) \right)\,,
\eeq
where $\Hhard$ is a finite hard-scattering amplitude 
while ${\bf Z}$ captures all singularities.  ${\bf Z}$ 
admits a renormalization group equation whose solution 
(in the minimal-subtraction scheme) can be written as a 
path-ordered exponential of the soft anomalous dimension:
\beq \label{RGsol}
{\bf Z} \left(\{p_i\},\mu, \as (\mu) \right) \, = \,  
{\cal P} \exp \left\{ -\int_0^{\mu} \frac{\Dd \lambda}{\lambda}\,
{\bf \Gamma} \left(\{p_i\},\lambda, \as(\lambda) \right) \right\}.
\eeq
The scale dependence of the soft anomalous 
dimension ${\bf \Gamma}\left(\{p_i\},\lambda, \as\right)$ 
for \emph{massless-parton} ($p_i^2=0$) scattering is 
both explicit and via the $4-2\eps$ dimensional coupling. 
In QCD (with $n_f$ light quark flavours) the latter obeys 
the renormalization group equation
\beq\label{betagamma}
\beta(\as,\eps) \equiv \frac{d\as}{d\ln \mu}= 
-2\eps \,\as - \frac{\as^2}{2\pi} \sum_{n = 0}^{\infty} 
b_n \, \left(\frac{\as}{\pi}\right)^n\,\qquad \text{with}\qquad  
b_0=\frac{11}{3}C_A-\frac{2}{3}\nf\,.
\eeq
For our purposes only the zeroth order solution will be needed: 
$\as(\mu) = \as(p) \left(p^2/\mu^2\right)^{\eps}$. The explicit 
dependence on the scale (${\bf \Gamma}$ is linear in 
$\log\lambda$) reflects the presence of double poles 
due to overlapping soft and collinear divergences.

The soft anomalous dimension in multileg scattering 
of massless partons is an operator in colour space 
given by~\cite{Becher:2009cu,Gardi:2009qi,Becher:2009qa,Gardi:2009zv,Almelid:2015jia}
\begin{eqnarray} 
\label{gammaSoft1}
&&\hspace{-0.3cm}{\bf \Gamma}\left(\{p_i\},\lambda, \as(\lambda) \right) \,=\,
{\bf \Gamma}^{\rm dip.}\left(\{p_i\},\lambda, \as(\lambda) \right)
\,+\,\sum_{n=3}^{\infty}{\bf \Delta}^{(n)}\left(\frac{\alpha_s}{\pi}\right)^n, \\ &&  \nn \\
&&\hspace{-0.5cm}\quad\text{with}\qquad \nn
{\bf \Gamma}^{\rm dip.}\left(\{p_i\},\lambda, \as(\lambda) \right)=
-\frac{\gamma_{K}  (\as)}{2} \, \, \sum_{i<j} 
\log \left(\frac{-s_{ij}}{\lambda^2}\right) \, \T_i \cdot \T_j \,+\, \sum_i \gamma_i (\as) \,,
\end{eqnarray}
where ${\bf \Gamma}^{\rm dip.}$ involves only 
pairwise interactions amongst the hard partons, 
and is therefore referred to as the ``dipole formula''.
The kinematic variables are $-s_{ij} = 2 |p_i \cdot p_j 
| e^{-i \pi \lambda_{ij}}$ with $\lambda_{ij} = 1$ if 
partons $i$ and $j$ both belong to either the initial 
or the final state and $\lambda_{ij} = 0$ otherwise.
The function ${\gamma}_{K}(\alpha_s)$ in \eqn{gammaSoft1} 
is the (lightlike) cusp anomalous dimension
\cite{Korchemsky:1985xj,Korchemsky:1985xu,Korchemsky:1987wg}, 
\emph{divided} by the quadratic Casimir of the corresponding 
Wilson lines.  
The functions $\gamma_i (\as)$ represent the field 
anomalous dimension corresponding to the parton~$i$, 
which governs hard collinear singularities. Both 
${\gamma}_{K}(\alpha_s)$ and $\gamma_i (\as)$ 
are known through three-loop in QCD and their 
values are summarised in Appendix A of 
ref.~\cite{Caron-Huot:2017fxr}. In \eqn{gammaSoft1}
$\Delta^{(n)}$ for $n\geq 3$ accounts for multi-parton 
correlations. The three-loop correction $\Delta^{(3)}$, 
correlating up to four hard partons, was calculated 
recently~\cite{Almelid:2015jia,Gardi:2016ttq} for any 
number of partons in general kinematics. Specializing 
to $2\to 2$ parton scattering in the high-energy limit, 
ref.~\cite{Caron-Huot:2017fxr} showed that $\Delta^{(3)}$
contributes starting from NNLL accuracy in the imaginary 
(even) part of the amplitude, and starting from N$^3$LL 
accuracy in the real (odd) part; we refer the interested 
reader to eq.~(4.11) in ref.~\cite{Caron-Huot:2017fxr} 
for an expression for $\Delta^{(3)}$ in this limit. Given 
our focus here on NLL accuracy, we shall not discuss 
it further.

While it is known that ${\bf \Gamma}^{\rm dip.}$ fully 
describes the infrared singularities associated with 
Regge pole factorisation \cite{Bret:2011xm,DelDuca:2011ae} 
--- meaning it is exact at leading and NLL accuracy for the 
real part the amplitude --- it does not fully capture the 
structure of the two-Reggeon cut~\cite{Caron-Huot:2013fea} 
at NLL accuracy, where $\Delta^{(n)}$ at four loops and 
beyond, are relevant. To identify the contribution of the 
soft anomalous dimension in two-parton scattering, 
$ij\to ij$, at increasing logarithmic accuracy, let 
us expand ${\bf \Gamma}$ in powers of $\alpha_s$, 
keeping the product $\alpha_s L$ fixed, as follows:
\be \label{eq:gammas}
{\bf \Gamma} \left(\as(\lambda)\right) = 
{\bf \Gamma}_{\LL}\left(\as(\lambda),L\right) 
+ {\bf \Gamma}_{\NLL}\left(\as(\lambda),L\right) 
+ {\bf \Gamma}_{\NNLL} \left(\as(\lambda),L\right) 
+ \ldots\,.
\ee
The N$^k$LL term in \eqn{eq:gammas} can be written 
as an expansion in $\alpha_s^m L^{m-k}$ for $m\geq 1$.
Using Regge-pole factorisation it can be 
shown~\cite{Bret:2011xm,DelDuca:2011ae}  
that the leading logarithmic contribution 
${\bf \Gamma}_{\LL}$ takes the one-loop 
exact form,
\be \label{eq:gammall}
{\bf \Gamma}_{\LL} \left(\as(\lambda)\right) \, =  \, 
\frac{\as(\lambda)}{\pi} \frac{\gamma^{(1)}_{K}}{2} L \, \Tt\,
\, =  \,  \frac{\as(\lambda)}{\pi}  L \, \Tt\,.
\ee
This exactly corresponds to the infrared-divergent 
part of the one-loop gluon Regge trajectory in 
\eqn{GluonRegge1}. 
Note that the LL anomalous dimension has even signature ${\bf \Gamma}_{\LL}={\bf \Gamma}_{\LL}^{(+)}$.
At NLL the anomalous dimension 
can be divided into signature-even and odd parts:
\be \label{GammaNLL0}
{\bf \Gamma}_{\NLL} = {\bf \Gamma}_{\NLL}^{(+)}  + {\bf \Gamma}_{\NLL}^{(-)}.
\ee
The even part\footnote{Note that the even part of the NLL 
anomalous dimension, ${\bf \Gamma}_{\NLL}^{(+)}$, 
contributes to the \emph{odd} NLL amplitude, 
${\cal M}_{\rm NLL}^{(-)}$, since it acts on the LL  
part of ${\cal H}$ in eq.~(\ref {IRfacteq}), which is itself odd.}, 
which is governed by the Regge pole, 
is two-loop exact. Referring to \eqn{gammaSoft1}, it 
contains the terms in the one-loop anomalous dimension 
that are not enhanced by $L$, as well as the infrared-divergent 
part of the two-loop gluon Regge trajectory:
\be \label{GammaNLL1b}
{\bf \Gamma}_{\NLL}^{(+)} =  \frac{\as(\lambda)}{\pi}
\sum_{i=1}^{2} \left( \frac{\gamma^{(1)}_{K}}{2} C_i \log \frac{-t}{\lambda^2} 
+ 2\gamma^{(1)}_i \right)\,+\, \left(\frac{\as(\lambda)}{\pi}\right)^2 
\frac{\gamma^{(2)}_{K}}{2} L \, \Tt.
\ee 
The odd part is however sensitive the the two-Reggeon 
cut. At one-loop it can be obtained from the dipole 
formula~\cite{Bret:2011xm,DelDuca:2011ae},
\be \label{GammaNLL1}
{\bf \Gamma}_{\NLL}^{(-)} =
 i  \pi \frac{\as(\lambda)}{\pi}\, \Tsu + O(\as^4L^3)\,,
\ee
while higher-order terms have so far been unknown.
The reduced amplitude obtained in section \ref{soft} 
contains information on the infrared divergences of 
next-to-leading high-energy logarithms to all orders in 
$\alpha_s$, and hence allows us to determine 
${\bf  \Gamma}_{\NLL}^{(-)}$ to all orders. 

In order to make contact with section \ref{soft} we 
need to express the reduced amplitude defined in 
\eqn{Mreduced} in its infrared-factorised form. Focusing 
on the even component, we substitute \eqn{IRfacteq} there 
and expand it to NLL accuracy:
\begin{align}\label{Mreduced-IR-1}
\begin{split}\Mreduced^{(+)}_{\NLL} =
\exp \bigg\{-\frac{\as(\mu)}{\pi}\frac{\bn{0}(\eps)}{2\eps} \, L \Tt\bigg\} \,\, 
\bigg[& {\bf Z}^{(-)}_{\NLL} \left(\frac{s}{t},\mu, \as (\mu) \right) \,
\Hhard^{(-)}_{\LL} \left(\{p_i\},\mu, \as (\mu) \right) \\&
+ {\bf Z}^{(+)}_{\LL} \left(\frac{s}{t},\mu, \as (\mu) \right) \,
\Hhard^{(+)}_{\NLL} \left(\{p_i\},\mu, \as (\mu) \right)  \bigg],
\end{split}
\end{align}
where we have written the Regge trajectory 
explicitly according to \eqn{GluonRegge1}. 
Substituting ${\bf \Gamma}_{\LL}$
of \eqn{eq:gammall} into \eqn{RGsol} and 
integrating over the scale (using the zeroth-order 
scale dependence of  $\as$) we obtain:
\be \label{ZLL}
 {\bf Z}^{(+)}_{\LL} \left(\frac{s}{t},\mu, \as (\mu) \right) = 
 \exp \bigg\{\frac{\as}{\pi} \frac{1}{2\eps} \, L \Tt\bigg\}\,.
\ee
Considering the second term in the square 
brackets of \eqn{Mreduced-IR-1} we note that 
${\bf Z}^{(+)}_{\LL}$ can be combined with the 
exponential of the Regge trajectory, and this 
combination gives rise to an exponent proportional 
to $(\bn{0}(\eps)-1)/(2\eps) \sim {\cal O}(\eps)$. 
Given that the hard function is finite by definition, 
$\Hhard^{(+)}_{\NLL} \sim {\cal O}(\eps^0)$, we 
conclude that the second term in \eqn{Mreduced-IR-1} 
only contributes to finite terms in $\Mreduced^{(+)}_{\NLL} $. 
This implies that the infrared-singular part of 
the reduced amplitude is insensitive to 
$\Hhard^{(+)}_{\NLL}$ \cite{Caron-Huot:2013fea} and is given by:
\be\label{Mreduced-IR-2}
\Mreduced^{(+)}_{\NLL} =
\exp \bigg\{-\frac{\as}{\pi} \frac{\bn{0}(\eps)}{2\eps} \, L \Tt\bigg\} \,
{\bf Z}^{(-)}_{\NLL} \left(\frac{s}{t},\mu, \as (\mu) \right) \,
\Hhard^{(-)}_{\LL} \left(\{p_i\},\mu, \as (\mu) \right) + \ord(\eps^0).
\ee

Equation~(\ref{Mreduced-IR-2}) can be further simplified 
by noticing that the hard function at LL accuracy 
is fixed by Regge factorisation: it is simply the 
exponential of the finite part of the gluon Regge 
trajectory, i.e.\ we have 
\be \label{HLL}
\Hhard^{(-)}_{\LL} \left(\{p_i\},\mu, \as (\mu) \right) 
=  \exp \bigg\{\frac{\as}{\pi} \frac{\bn{0}(\eps)-1}{2\eps} 
\, L \CA\bigg\} \, \Mtree, 
\ee
where we used the fact that $\Tt = C_A$ when 
acting on the Regge limit of the tree level amplitude.
Moving this (finite) exponential to the left, this result 
allows us to write \eqn{Mreduced-IR-2} more 
explicitly as
\begin{multline} \label{eq:mhatsoft0}
\exp\left\{\frac{(1-\bzero(\eps) )}{2\eps} \frac{\as}{\pi} L (\CA-\Tt) \right\} 
\Mreduced_\NLL = \exp\left\{-\frac{1}{2\eps} \frac{\as}{\pi} L \Tt \right\}
\\ \times
\po \exp\left\{-\int_0^{p} \frac{\Dd \lambda}{\lambda} 
\bigg[ {\bf \Gamma}_{\LL} \left(\as(\lambda)\right) 
+ {\bf \Gamma}_{\NLL} \left(\as(\lambda)\right) \bigg]\right\} 
\, \Mtree + \ord(\eps^0),
\end{multline}
where it is understood that both sides of this equality 
are to be projected onto even signature.
Below we will abbreviate the l.h.s.\ as $\Mbar_\NLL$.
The NLL contribution to the path-ordered 
exponential on the second line can be 
written out fully as
\be
-\int_0^p \frac{d\lambda}{\lambda}
\bigg[\po \exp\left\{-\int_0^{\lambda} 
\frac{d\lambda'}{\lambda'}{\bf \Gamma}_{\LL} 
\left(\as(\lambda')\right)\right\}\bigg]
{\bf \Gamma}_{\NLL} \left(\as(\lambda)\right)
\bigg[\po \exp\left\{-\int_{\lambda}^p 
\frac{d\lambda'}{\lambda'}{\bf \Gamma}_{\LL} 
\left(\as(\lambda')\right)\right\}\bigg]\,.
\ee
Finally, integrating the exponents in each of the two brackets as in \eqn{ZLL} 
and using again that $\Tt= C_A$ in the right factor upon acting on $\Mtree$, 
we obtain, projecting onto the even amplitude:
\begin{align} \label{eq:mhatsoft}
\Mbar_\NLL^{(+)} = -\int_0^p \frac{d\lambda}{\lambda}
\exp\left\{\frac{1}{2\eps} \frac{\as(p)}{\pi} L (\CA-\Tt) 
\left[1-\mratio{p}{\lambda}^{\!\eps}\,\right]\right\} 
{\bf \Gamma}_{\NLL}^{(-)} \left(\as(\lambda)\right) 
\, \Mtree + \ord(\eps^0).
\end{align}
This expression for the even amplitude may be 
compared directly with the one obtained in 
\eqn{eq:mredregge} using the BFKL analysis; 
exploiting the fact that the exponential on the 
l.h.s.\ of \eqn{eq:mhatsoft0} is finite (and that 
$R(\eps)$ is finite), the BFKL prediction can 
be written as
\begin{align} \label{eq:mhatreggeB}
\Mbar_\NLL^{(+)}= i\pi
  \left[\frac{\displaystyle\exp\left\{\frac{1}{2\eps} 
  \frac{\as}{\pi} L (\CA-\Tt)\right\} - 1}{L(\CA-\Tt)}\right]
\left( 1 - \frac{C_A}{C_A -\Tt} R(\eps) \right)^{-1}
   \Tsu \, \Mtree + \ord(\eps^0)
\end{align}
with $R(\eps)$ defined in \eqn{Rdef}.
We now have two expressions for the infrared 
singularities of the reduced amplitude ---
an expression in terms of the 
soft anomalous dimension, \eqn{eq:mhatsoft},
and the all-order  result of BFKL evolution 
in the soft approximation, \eqn{eq:mhatreggeB}. 
In the next section we equate them and extract 
${\bf \Gamma}^{(-)}_{\NLL}$.

\subsection{Extracting the soft anomalous dimension at NLL}
\label{ssec:extractgamma}

In minimal subtraction schemes, anomalous 
dimensions can be extracted by taking
the coefficient of pure $1/\eps$ single poles.
Indeed, to get the coefficient of the single 
poles in \eqn{eq:mhatsoft} we can drop the 
exponentials to get
\bea\nn
\Big[\Mbar_\NLL^{(+)}\Big]_{\text{single poles}} &=& 
-\int_0^{p} \frac{\Dd \lambda}{\lambda} \, 
{\bf \Gamma}_{\NLL}^{(-)} \left(\as(\lambda)\right)\, \Mtree
\\&=&
\frac{1}{2\eps} \sum_{\ell =1}^{\infty}  
\left(\frac{\as(p)}{\pi}\right)^{\ell}\, L^{\ell-1}\, \frac{1}{\ell}\,
{\bf \Gamma}^{(-,\ell)}_{\NLL}
\, \Mtree\,.
\label{eq:mhatsoft-singlepoles}
\eea
This result must be set equal to the 
single poles obtained from \eqn{eq:mhatreggeB},
whose $\ell$-loop coefficient is
\be
\Mbar_\NLL^{(+,\ell)}= 
\frac{i\pi}{2\eps\,\ell!} \left[ \frac{(\CA-\Tt)}{2\eps}\right]^{\ell-1} 
\left( 1 -  \frac{C_A}{C_A -\Tt} R(\eps)\right)^{-1} \Tsu \, \Mtree + \ord(\eps^0).
\ee
Comparing with \eqn{eq:mhatsoft-singlepoles} 
then gives
\be
\label{G_NLL_Gl}
{\bf \Gamma}^{(-,\ell)}_{\NLL} = i\pi \,G^{(\ell)} \,\Tsu
\ee
with
\be
\label{Gl}
G^{(\ell)} \equiv \frac{1}{(\ell-1)!}\left[ \frac{(\CA-\Tt)}{2}\right]^{\ell-1} 
\left.\left( 1 - \frac{C_A}{C_A -\Tt}R(\eps)  \right)^{-1}\right\vert_{\eps^{\ell-1}} \,\,,
\ee
where the subscript indicates that one should
extract the coefficient of $\eps^{\ell-1}$. 
Although the notation does not manifest 
this, the end result is always a polynomial 
in colour operators $\CA$ and $\Tt$, since 
$R(\eps)$ has a regular series as $\eps\to 0$.
Rescaling $\eps$, this can also be written as
\be
\label{main_result}
{\bf \Gamma}^{(-,\ell)}_{\NLL}
= \frac{i\pi}{(\ell-1)!} 
\left.\Bigg(1 -  \frac{\CA}{\CA-\Tt}\R\left(x{(\CA-\Tt)}/{2}\right) \Bigg)^{-1}\right\vert_{x^{\ell-1}}
\,\Tsu\,.
\ee
where the function $R(\eps)=-2\zeta_3\,\eps^3+\ldots$ 
is defined in \eqn{Rdef}.

Equation~(\ref{main_result}) is the main 
result of this paper: it gives the soft anomalous dimension 
in the Regge limit to any loop order at next-to-leading 
logarithmic accuracy (i.e.\ all terms of the form 
$\alpha_s^\ell L^{\ell-1}$); the even contribution 
${\bf \Gamma}^{(+,\ell)}_{\NLL}$ was given in 
\eqns{eq:gammall}{GammaNLL1b}. In other words, 
we now know \eqn{GammaNLL1} to all orders:
\be \label{GammaNLL2}
{\bf \Gamma}_{\NLL}^{(-)} =
\sum_{\ell=1}^{\infty} {\bf \Gamma}_{\NLL}^{(-,\ell)}
\left(\frac{\alpha_s(\lambda)}{\pi}\right)^{\ell} L^{\ell-1} \,.
\ee
Expanding the above formula explicitly to eight loops:
\begin{align}
\label{eq:gamma8}
\begin{split}
{\bf \Gamma}^{(-,1)}_{\rm NLL} &= 
i \pi \, \Tsu  \\
{\bf \Gamma}^{(-,2)}_{\rm NLL} &= 
0   \\
{\bf \Gamma}^{(-,3)}_{\rm NLL} &= 
0,  \\
{\bf \Gamma}^{(-,4)}_{\rm NLL} &= 
- i \pi \, \frac{\zeta_3}{24}\,C_A (C_A - \Tt)^2   \, \Tsu,  \\
{\bf \Gamma}^{(-,5)}_{\rm NLL} &= 
- i \pi \, \frac{\zeta_4}{128}\,C_A (C_A - \Tt)^3   \, \Tsu, \\
{\bf \Gamma}^{(-,6)}_{\rm NLL} &= 
- i \pi \, \frac{\zeta_5}{640}\,C_A (C_A - \Tt)^4   \, \Tsu, \\ 
{\bf \Gamma}^{(-,7)}_{\rm NLL} &= 
i \pi \frac{1}{720} \bigg[\frac{\zeta_3^2}{16}\,C_A^2 (C_A - \Tt)^4
+\frac{1}{32} \left(\zeta_3^2 - 5 \zeta_6\right) \,C_A (C_A - \Tt)^5  \bigg] \Tsu,  \\ 
{\bf \Gamma}^{(-,8)}_{\rm NLL} &= 
i \pi \frac{1}{5040} \bigg[\frac{3 \zeta_3\zeta_4}{32}\,C_A^2 (C_A - \Tt)^5
+\frac{3}{64} \left(\zeta_3 \zeta_4 - 3 \zeta_7\right) \,C_A (C_A - \Tt)^6  \bigg] \Tsu.  
\end{split}
\end{align}
These results are valid in any gauge theory, and 
hold modulo colour operators which vanish when 
acting on the Regge limit of the tree amplitude
(which is given by the $t$-channel 
gluon exchange diagram).

\subsection{Properties of the soft anomalous dimension in the Regge limit}

\def\singlet{{1}}
\def\twentyseven{{27}}

In the previous section we computed 
${\bf \Gamma}^{(-)}_{\NLL}$, the imaginary 
part of the soft anomalous dimension in the 
Regge limit,  to all orders. Let us briefly 
explore its properties addressing the 
colour structure, the convergence of the 
expansion, and finally its asymptotic 
high-energy behaviour.

Considering eq.~(\ref{eq:gamma8}), our first 
observation is that colour structures of 
increasing complexity emerge every three 
loops, as dictated by the expansion of $R(\epsilon)$ 
in \eqn{Rdef}: corrections going beyond the 
dipole formula start at four loops, where the 
colour structure is proportional to $C_A$ to 
a single power. This correction reproduces 
precisely that found previously in ref.~\cite{Caron-Huot:2013fea}.
Proceeding to five and six loops ${\bf\Gamma}_{\rm NLL}$ 
only incurs extra powers of $(C_A - \Tt)$. Starting 
at seven loops, however terms with two powers of 
$C_A$ appear as well. Similarly, a cubic power of 
$C_A$ would emerge at ten loops, and so on.
We also note that the zeta values appearing in 
${\bf\Gamma}_{\rm NLL}$ are of uniform weight,
which is, of course, again a mere consequence 
of the Taylor series of $R(\epsilon)$.

To proceed it would be useful to specify the relevant 
colour charge exchanged in the $t$ channel, $\Tt$. 
To this end consider for example gluon-gluon scattering, 
where the $t$ channel colour flow can be any of the 
${\rm SU}(N_c)$ representations appearing in the 
decomposition\footnote{A more complete exposition 
of the $t$-channel basis of colour flow can be found 
in refs.~\cite{DelDuca:2014cya,Caron-Huot:2017fxr}.} 
\begin{equation}
8 \otimes 8 = \singlet \oplus 8_{s} \oplus 8_{a} 
\oplus 10 \oplus \overline{10} \oplus \twentyseven \oplus 0\,,
\end{equation}
where the labels refer to their dimensions for $N_c=3$.
Because of Bose symmetry, 
the symmetry of the colour structure mirrors the 
signature of the corresponding amplitudes under 
$s \leftrightarrow u$  exchange. Thus, only even 
representations are relevant for the two-Reggeon 
amplitude discussed here; these are the singlet, 
where $\Tt=0$, the symmetric octet with $\Tt=C_A=N_c$, 
the $\twentyseven$ representation with $\Tt= 2 (N_c + 1)$, 
and the ``0'' representation, where $\Tt= 2 (N_c - 1)$. In 
the following we restrict the discussion to the first three 
cases, which are all relevant for QCD with $N_c=3$ 
(the latter has a  vanishing dimension, and hence it 
does not contribute). 

The next observation, already mentioned in 
section~\ref{multiloop-exact}, is that the symmetric 
octet representation with $\Tt=C_A$, corresponds 
to a constant wavefunction, and thus a trivial solution 
to eq.~(\ref{Hdef1}), with no corrections to the reduced 
amplitude beyond one loop (as can be verified for 
example in the explicit results in 
eqs.~(\ref{MellReggeSoft123}) through 
(\ref{MellReggeSoft101112}) upon considering  $\Tt=C_A$). 
The reduced amplitude for the symmetric octet state is thus 
one-loop exact, corresponding to a simple Regge-pole 
behaviour with a gluon Regge trajectory for the original 
amplitude according to eq.~(\ref{Mreduced}). This of 
course reproduces the known behaviour of the 
symmetric-octet exchange used in the original 
derivation of the BFKL equation. 
In turn, for the singlet --- the famous Pomeron --- and 
$\twentyseven$ representation, we find non-trivial 
radiative corrections associated with a Regge cut. 
We will thus use these two examples in the 
discussion that follows. 

\begin{figure}[t]
  \centering
  \includegraphics{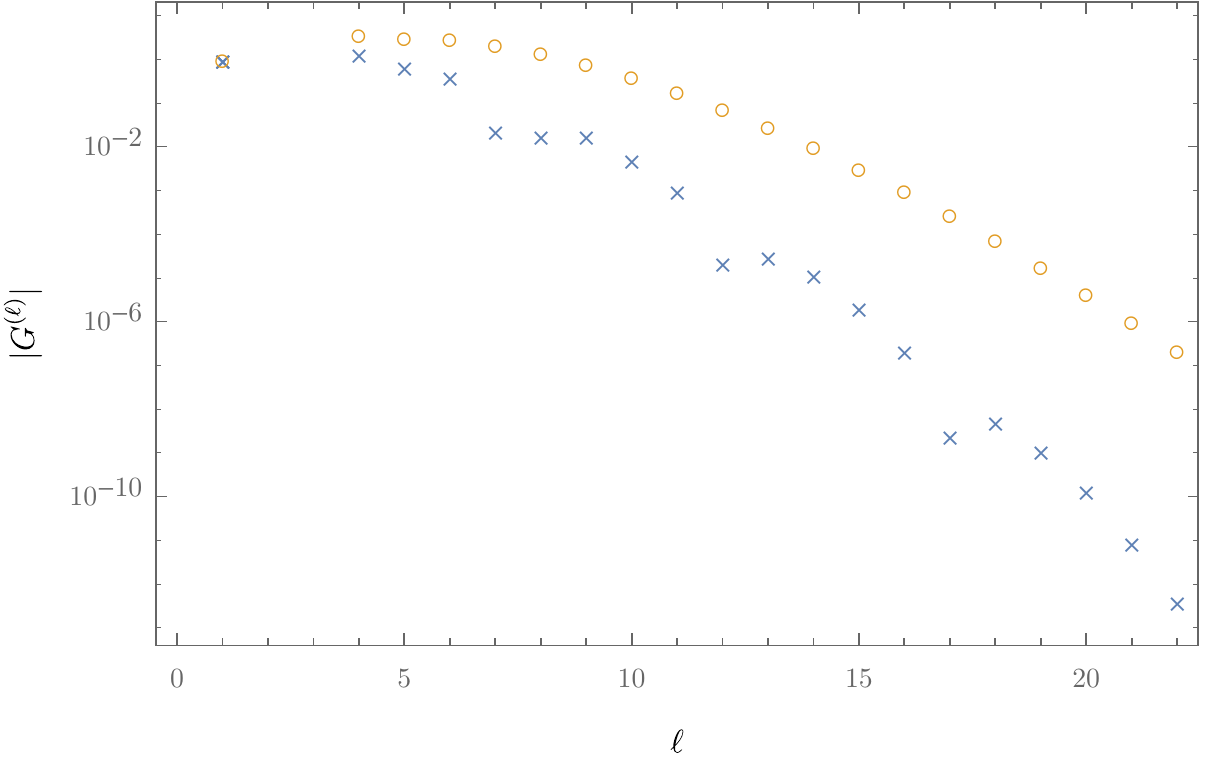}
  \caption{Logarithmic plot of the absolute value of the 
  coefficients $G^{(\ell)}$ \eqref{Gx}, for $\ell=1,\dots,22$. 
  The $|G^{(\ell)}|$ quickly become very small suggesting 
  good convergence of the series. Shown is the singlet 
  (crosses) and $\twentyseven$ exchange (circles).}
  \label{fig-coef_plot0_27}
\end{figure}

Next let us consider the convergence properties 
of the perturbative series representing the soft 
anomalous dimension in eq.~(\ref{G_NLL_Gl}). 
One immediately notes that this series is highly 
convergent due to the $1/(\ell-1)!$ prefactor in 
eq.~(\ref{Gl}). Figure~\ref{fig-coef_plot0_27} 
illustrates this factorial suppression of the 
coefficients $G^{(\ell)}$ as a function of the 
order $\ell$ for $C_A=N_c=3$ and for the two 
relevant representations, the singlet and the 
$\twentyseven$.

Furthermore, we can establish that the anomalous dimension 
(\ref{main_result}) has \emph{an infinite radius of convergence}
as a function of $x\equiv L\as/\pi$.
To see this we write the resummed 
soft anomalous dimension as:
\be \label{GammaNLL3}
{\bf \Gamma}_{\NLL}^{(-)} = i\pi \frac{\alpha_s}{\pi} 
\,G\left(\frac{\alpha_s}{\pi}L\right)\Tsu\,,
\ee
where the generating function for the expansion 
coefficients is defined by
\be
\label{Gx}
 G(x) = \sum_{\ell=1}^\infty x^{\ell-1} G^{(\ell)}\,.
\ee
It is convenient to further identify $G(x)$ as the  
Borel transform of some function 
\be
\label{gGborel}
g(y)\equiv \int_0^{\infty}\! dx  \,G(x) \,e^{-x/y}
=\sum_{\ell=1}^\infty G^{(\ell)} y^\ell (\ell-1)!\,,
\ee
which upon using eq.~(\ref{Gl}), simply evaluates to
\be
\label{gy}
g(y)=\frac{y}{1 - {\displaystyle\frac{\CA}{\CA-\Tt}\,\R\left(y{(\CA-\Tt)}/{2}\right)}}\,.
\ee
We may now recover the original $G(x)$ via the integral
\be \label{Ggrelation}
  G(x) = \frac{1}{2\pi i}\int_{w-i\infty}^{w+i\infty} 
  d\eta\, g\left(\frac{1}{\eta}\right) {\rm e}^{\eta x}\,,
\ee
where the integration contour 
runs parallel to the imaginary axis, 
to the right of all singularities of the integrand. 
\begin{figure}[t]
  \centering
  \includegraphics{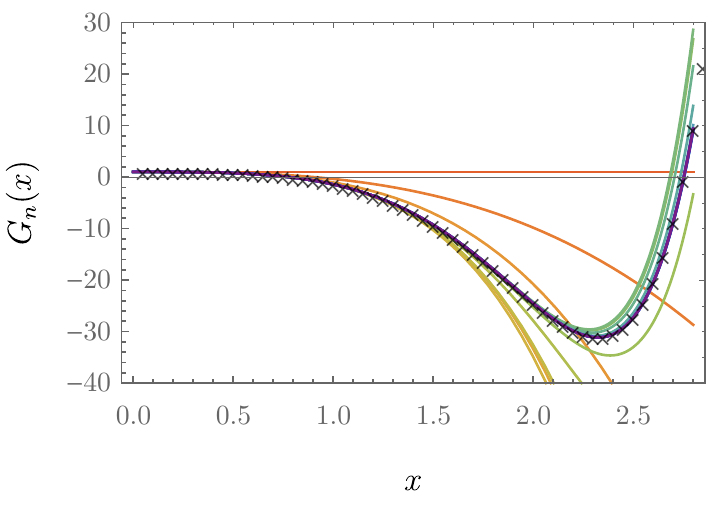}\hspace{9pt} 
  \includegraphics{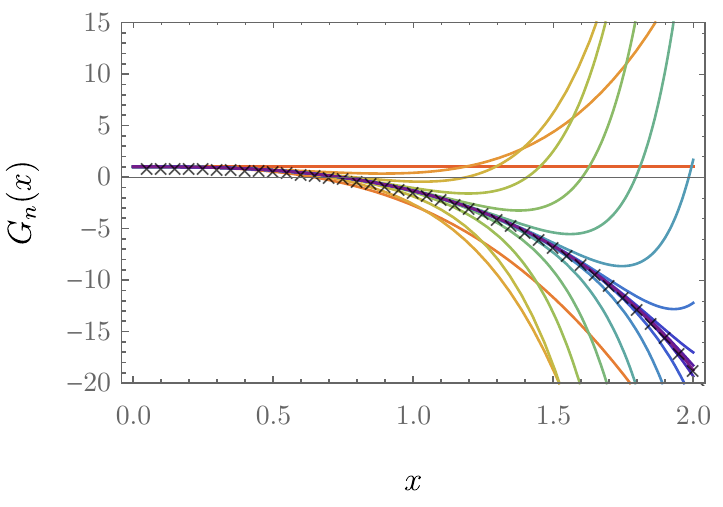}
  \caption{Partial sums $G_n(x) = \sum_{\ell=1}^{n} 
  G^{(\ell)} x^{\ell-1}$ for $n=1,\dots,22$ (rainbow, red 
  through violet) and numerical results for $G(x)$ (black 
  crosses). The plot illustrates convergence in that 
increasing the order~$n$ extends the range of $x$ for which the the partial 
  sum matches the numerical result. The figure shows 
  the singlet (left) as well as the $\twentyseven$ 
  exchange (right).}
  \label{fig:Glsums}
\end{figure}

The function $g(y)$ in eq.~(\ref{gGborel}) only 
has isolated poles away from the origin and has 
a finite radius of convergence: it is well-defined 
in a disc around the origin. It then follows that 
$G(x)$ has an infinite radius of convergence, 
hence this function --- and the soft anomalous 
dimension ${\bf \Gamma}_{\NLL}^{(-)}$ in 
\eqn{GammaNLL3} --- is \emph{an entire function}, 
free of any singularities for any finite $x = L \as /\pi$.

We stress that our use of the Borel transform is 
opposite to the usual application of Borel summation 
(which is ordinarily used to sum asymptotic series): 
the function $G(x)$, in which we are interested, is an entire 
function; we make use of its \emph{inverse} Borel 
transform, $g(y)$, which has \emph{worse} behaviour 
by having merely a finite radius of convergence. 
Nonetheless we find that 
numerically integrating eq.~(\ref{Ggrelation}) is a particularly 
convenient way to evaluate the anomalous dimension.
This numerical integration is compared to the partial 
sums 
\begin{equation}
\label{parsum}
G_n(x) \equiv \sum_{\ell=1}^{n} G^{(\ell)} x^{\ell-1}
\end{equation}
in figure~\ref{fig:Glsums}, where we find good agreement 
for the given values of $x$. While it becomes 
challenging to efficiently compute the coefficients 
$G^{(\ell)}$ at high orders (here we only evaluated them 
for $\ell \leq 22$), we find the numerical integration of 
eq.~(\ref{Ggrelation}) to be very stable, even for larger 
values of $x$. Thus, the remarkable convergence 
properties of $G(x)$ along with the Borel technique, 
presents us with the possibility of computing 
$\Gamma_{\rm NLL}^{(-)}$ for $x=L \alpha_s /{\pi} \,\gg\,1$, 
i.e. at asymptotically high energies. 
This is a rather unique situation in a perturbative 
setting --- in other circumstances resummation 
techniques are limited to the region 
$x=L {\alpha_s} /{\pi}\,\lesssim\,1$.

\begin{figure}[t]
  \centering
  \includegraphics{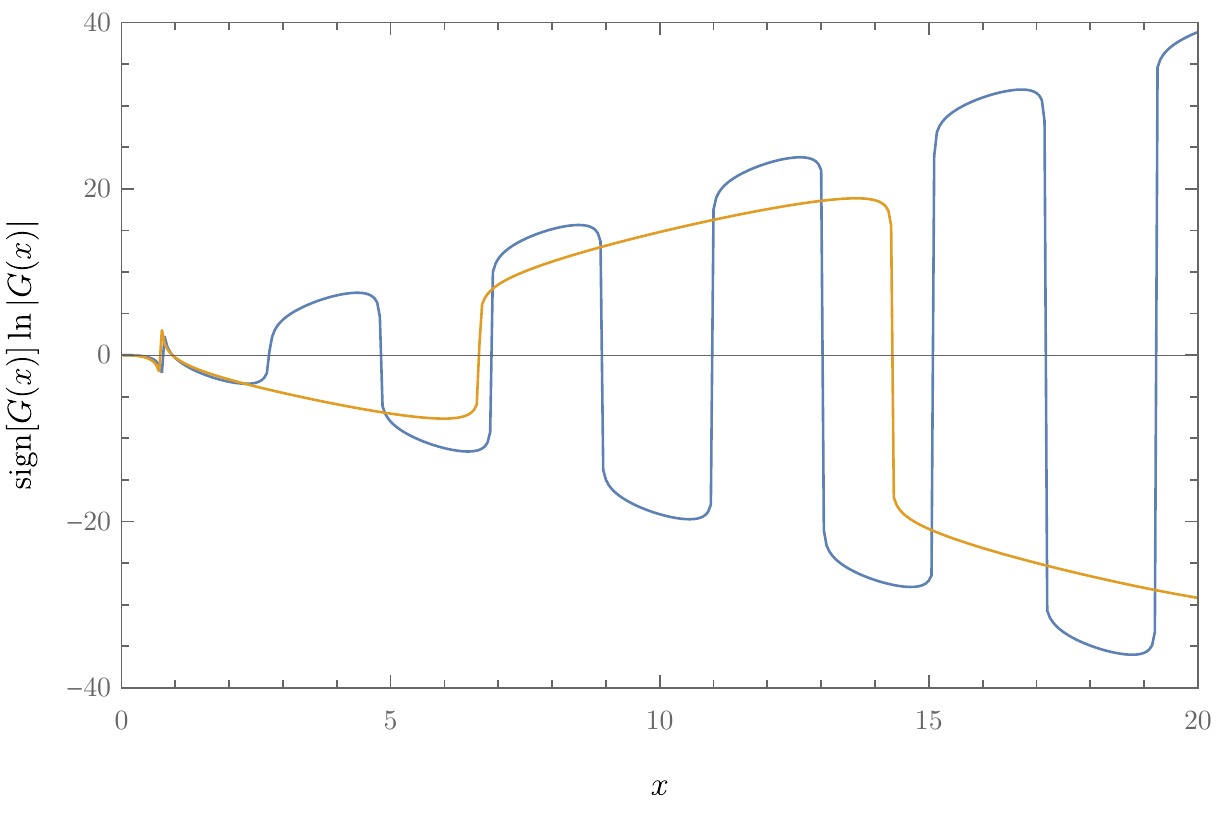}
  \caption{Numerical results for $\mathrm{sign} 
  \left[ G(x) \right] \ln \left| G(x) \right|$ for the singlet (blue) 
  and $\twentyseven$ exchange (orange). The ``heartbeat'' 
  at small $x$ reflects the logarithmic divergence of 
  $\ln \left| G(x) \right|$ where $G(x)$ changes its sign 
  for the first time (similar divergences occur every 
  oscillation but are not visible due to the finite 
  resolution of the plot).}
  \label{fig:Gofx}
\end{figure}

\begin{figure}[t]
  \centering
  \includegraphics{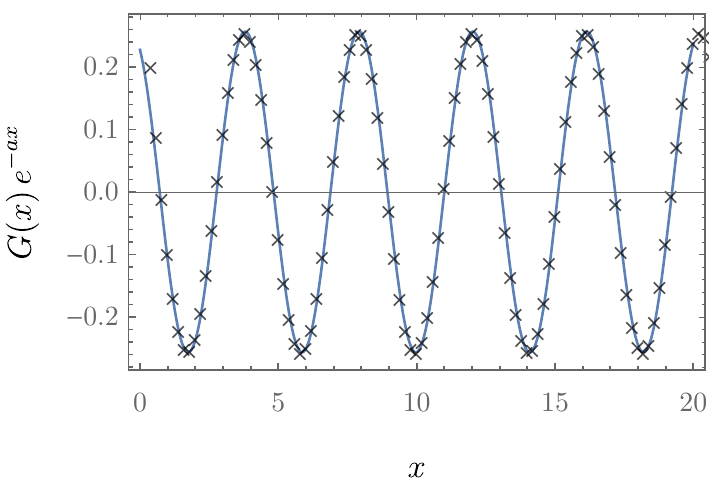}\hspace{9pt} 
  \includegraphics{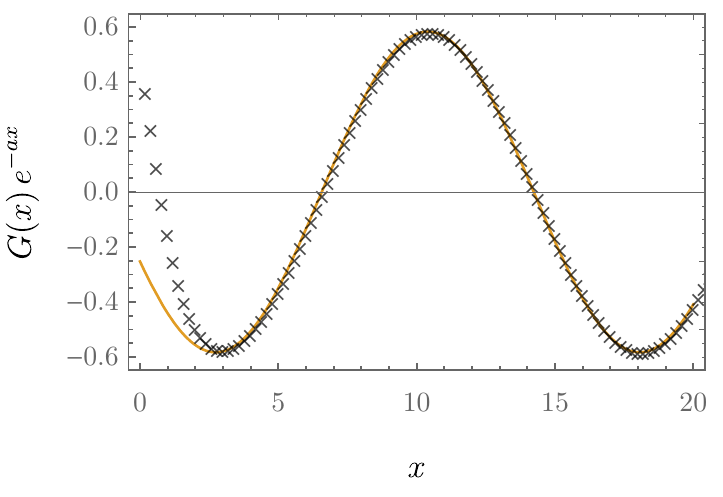}
  \caption{The approximation of eq.~\eqref{eq:Gmodel} for $G(x)$ for $x\gg 1$,
  divided by $e^{ax}$ (solid line) contrasted with numerical 
  results (crosses). The coefficients $a$ and $b$ were 
  extracted from the poles of $g(1/\eta)$ while $c$ and 
  $d$ were fitted after dividing the full, numerically evaluated, $G(x)$ by $e^{ax}$. Already for 
  moderate values of $x$ we observe excellent agreement. 
  The singlet exchange is shown on the left and the 
  $\twentyseven$ is on the right.}
  \label{fig:Gbyexp}
\end{figure}

Evaluating the integral (\ref{Ggrelation}) and 
plotting $G(x)$ for larger values of~$x$ reveals 
oscillations with a constant period and an exponentially 
growing amplitude. Since this behaviour is difficult to 
capture graphically we instead show the logarithm of 
$|G(x)|$ weighted by the sign of $G(x)$ in figure~\ref{fig:Gofx}.
This observation suggests to approximate (\ref{Ggrelation}) by
\be
  G(x) \to c\, e^{ax} \cos \left(bx + d\right)\,,
  \label{eq:Gmodel}
\ee
for sufficiently large values of $x$.  By 
means of \eqn{gGborel}, this model is equivalent to
\be
\label{gy_model}
g\left(\frac{1}{\eta}\right) \to  c\, \RE \left[ \frac{e^{id}}{\eta - a - ib} \right] 
= \frac{c}{2}\left( \frac{e^{id}}{\eta - a - ib} + \frac{e^{-id}}{\eta - a + ib} \right)\,,
\ee
which is to be integrated as in (\ref{Ggrelation}) with a contour to the right of the poles. 
\begin{table}[b]
  \centering
  \begin{tabular}{|c|c|c|c|c|}
  \hline
  & $a$ & $b$ & $c$ & $d$ \\
  \hline
  $\singlet$ & 1.97 & 1.52 & 0.25 & 0.48 \\
  $\twentyseven$ & 1.46 & 0.41 & 0.58 & 2.01\\
  \hline
  \end{tabular}
  \caption{Numerical results for $a,b,c$ and $d$, cf.\ 
  \eqn{eq:Gmodel}, for the singlet ($\singlet$) 
  and $\twentyseven$ representation.}
  \label{tab:abcdnumerical}
\end{table}
We thus find that to capture the behaviour $G(x)$ at 
large $x$ it is sufficient to simply consider 
$g\left(\frac{1}{\eta}\right)$ as a pair of 
complex-conjugated poles at $\eta = a \pm ib$. 
Indeed, numerically extracting the rightmost 
poles of $g\left(\frac{1}{\eta}\right)$ of 
eq.~(\ref{gy}) to identify the parameters 
$a$ and $b$ in eq.~(\ref{gy_model}), 
and dividing the full, numerically-evaluated, 
$G(x)$ by $e^{ax}$ leaves us with almost 
pure cosine-like behaviour for any $x\gg1$, 
as can be seen in figure~\ref{fig:Gbyexp}.
For reference, we quote our numerical results 
for $a,b,c$ and $d$ in table~\ref{tab:abcdnumerical}.

\subsection{Exponentiation check for higher-order infrared poles}
\label{ssec:higherpoles}

As a final step we confirm the agreement between 
the BFKL prediction and the soft  factorisation theorem. 
Thus far we have only used the single poles as predicted 
by the BFKL evolution to extract the NLL soft anomalous 
dimension ${\bf \Gamma}_{\NLL}^{(-)}$. As explained in 
section \ref{sec:IR_fact}, higher-order poles 
of the amplitude are generated upon expansion of 
the path-ordered exponential in \eqn{eq:mhatsoft}. 
They have to match the BFKL computation and therefore 
provide an independent and non-trivial check of our results. 

To see how this works, let us expand the BFKL result 
(\ref{eq:mhatreggeB}) to the first few orders, namely
\be\label{MbarEven}
\Mbar_{\rm NLL}^{(+)}\left(\frac{s}{-t}\right) 
= \sum_{\ell=1}^\infty \left( \frac{\as}{\pi} \right)^\ell
\,L^{\ell -1}\, \Mbar_{\rm NLL}^{(+,\ell)}\,.
\ee
with
\begin{subequations} \label{MbarNLL} \bea
\label{MbarNLL1} 
\Mbar_\NLL^{(+,1)} &=& i\pi \left[ \frac{1}{2\eps} 
+O(\eps^0)\right] \Tsu \Mtree,\\ 
\label{MbarNLL2} 
\Mbar_\NLL^{(+,2)} &=& i\pi \frac{(\CA-\Tt)}{2!}\left[ \frac{1}{(2\eps)^2} 
+O(\eps^0)\right] \Tsu \Mtree,\\ 
\label{MbarNLL3} 
\Mbar_\NLL^{(+,3)} &=& i\pi \frac{(\CA-\Tt)^2}{3!} \left[ \frac{1}{(2\eps)^3} 
+O(\eps^0)\right] \Tsu \Mtree,\\
\label{MbarNLL4} 
\Mbar_\NLL^{(+,4)} &=& i\pi \frac{(\CA-\Tt)^3}{4!} \left[ \frac{1}{(2\eps)^4} 
-\frac{1}{2\eps}\frac{\zeta_3 \CA}{4 (\CA-\Tt)}  +O(\eps^0)\right] \Tsu \Mtree,\\ \nn
\label{MbarNLL5}
\Mbar_\NLL^{(+,5)} &=& 
i\pi \frac{(\CA-\Tt)^4}{5!} \bigg[ \frac{1}{(2\eps)^5}
-\frac{1}{(2\eps)^2}\frac{\zeta_3 \CA}{4(\CA-\Tt)} \\ 
&&\hspace{3.0cm}-\, \frac{1}{2\eps}\frac{3\zeta_4 \CA}{16(\CA-\Tt)}
+O(\eps^0)\bigg] \Tsu \Mtree.
\eea \end{subequations} 
Let us begin with the leading pole. One can see a 
simple pattern in its $\ell$-th order coefficient, which is 
proportional to $(\CA-\Tt)^{\ell-1}/(\ell!(2\eps)^\ell)$. 
This should be compared with the prediction 
(\ref{eq:mhatsoft}) from infrared exponentiation, 
which we reproduce here for convenience:
\be \label{eq:mhatsoft-C}
\Mbar_\NLL^{(+)} = -\int_0^p \frac{d\lambda}{\lambda}
\exp\left\{\frac{1}{2\eps} \frac{\as(p)}{\pi} L (\CA-\Tt) 
\left[1-\mratio{p}{\lambda}^{\eps}\,\right]\right\} 
{\bf \Gamma}_{\NLL}^{(-)} \left(\as(\lambda)\right) \, \Mtree + \ord(\eps^0).
\ee
Substituting ${\bf \Gamma}_{\NLL}^{(-)}$ using
eqs.~(\ref{GammaNLL2}) and (\ref{G_NLL_Gl}), 
and taking into account that the running coupling 
$\as(\mu) = \as(p) \left(p^2/\mu^2\right)^{\eps}$, 
one gets 
\bea \label{eq:mhatsoft-C_subs}
\Mbar_\NLL^{(+)} =&& -i\pi\sum_{k=1}^{\infty} 
G^{(k)}
\left(\frac{\as(p)}{\pi}\right)^{k}\,L^{k-1} 
\int_0^p \frac{d\lambda}{\lambda}\mratio{p}{\lambda}^{\eps k}  \\ 
&& \nn
\qquad \times \exp\left\{\frac{1}{2\eps} \frac{\as(p)}{\pi} 
L (\CA-\Tt) \left[1-\mratio{p}{\lambda}^{\eps}\,\right]\right\} \,
\,\Tsu\, \Mtree + \ord(\eps^0).
\eea
For the leading pole it is clear that only the $G^{(1)}$ 
terms contribute, corresponding to the one-loop 
contribution to the soft anomalous dimension 
(\ref{GammaNLL1}), and we then get:
\bea
\nn
\left[ \Mbar_\NLL^{(+)} \right]_{\text{leading poles}} &=& 
-i\pi \, \frac{\as(p)}{\pi} \int_0^p \frac{d\lambda}{\lambda} 
\mratio{p}{\lambda}^{\eps}\\& \nn &
\quad\times\exp\left\{\frac{1}{2\eps} \frac{\as(p)}{\pi} 
L (\CA-\Tt) \left[1-\mratio{p}{\lambda}^{\eps}\,\right]\right\} 
\Tsu\Mtree
\\ &=& -i\pi \left[\frac{\displaystyle\exp\left\{\frac{1}{2\eps} 
\frac{\as(p)}{\pi} L (\CA-\Tt)\right\} - 1}{L(\CA-\Tt)}\right] \Tsu\Mtree\,.
\eea
Expanding in $\as$ this matches precisely the 
$1/(\ell!(2\eps)^\ell)$ terms in \eqn{MbarNLL}, 
with the correct prefactor. This exponentiation 
of leading poles had been verified previously in 
ref.~\cite{Caron-Huot:2013fea}. Moving on to 
the first subleading pole, the Regge prediction 
reveals a four-loop single pole in \eqn{MbarNLL4}, 
as well as a five-loop double pole in \eqn{MbarNLL5} 
and so on, all proportional to $\zeta_3$. In general, 
expanding the BFKL result (\ref{eq:mhatreggeB}) 
to higher orders one finds a tower of such terms
going like $1/(\ell!(2\eps)^{\ell-3})$.  In the infrared 
exponentiation formula, these should be generated 
by a single parameter, the four-loop anomalous 
dimension, ${\bf \Gamma}^{(-,4)}_{\rm NLL}$, 
which is indeed proportional to $\zeta_3$ (see 
\eqn{eq:gamma8}). It can be traced back to the 
leading-order term in the expansion of $R(\epsilon)$ 
in~(\ref{Rdef}), contributing to $G^{(4)}$ in \eqn{Gl}.
Similarly, a $k$-loop anomalous dimension 
${\bf \Gamma}^{(-,k)}_{\rm NLL}$, in general, 
contributes in proportion to $G^{(k)}$. Indeed, 
integrating \eqn{eq:mhatsoft-C_subs} we find that 
\begin{align} \label{eq:mhatsoft-C_subs_integrate}
\Mbar_\NLL^{(+)} =\frac{i\pi}{2\epsilon}
\sum_{k=1}^{\infty}\,G^{(k)}\, (k-1)!
\sum_{\ell=k}^{\infty} \frac{1}{\ell !}
\left(\frac{\as(p)}{\pi}\right)^{\ell}\,L^{\ell-1} 
\left(\frac{C_A-\Tt}{2\epsilon}\right)^{\ell-k}
\,\,\Tsu\, \Mtree + \ord(\eps^0).
\end{align}
Next we note that given $k$, all contributions 
with $\ell<k$ are either constant or vanish for $\epsilon\to 0$, 
and so in as far as the singularities are concerned the sum 
over $\ell$ can be performed over all positive integers, 
independently of $k$. This yields
\begin{align} \label{eq:mhatsoft-C_subs_integrate_summed}
\Mbar_\NLL^{(+)} =i\pi\, 
\sum_{k=1}^{\infty}\,\frac{G^{(k)}\, (k-1)! (2\epsilon)^{k-1}}{L\,\left({C_A-\Tt}\right)^{k}}\,
\left[{\displaystyle\exp\left\{\frac{1}{2\eps}  \frac{\as}{\pi} L (\CA-\Tt)\right\} - 1}\right]
\,\Tsu\, \Mtree + \ord(\eps^0).
\end{align}
This shows that infrared exponentiation works out 
\emph{if}, and \emph{only if}, all the poles in the NLL 
amplitude can be written as a function of $\eps$ only 
(i.e.\ independent of $\as$), times the quantity in the 
square bracket. With hindsight, infrared exponentiation 
thus explains the compact form of the BFKL result in 
\eqn{eq:mhatreggeB}. Finally, it is straightforward to 
substitute in the definition of $G^{(k)}$ from \eqn{Gl} 
and sum up the series over $k$, recovering the full 
result for the singularities of the amplitudes in 
\eqn{eq:mhatreggeB}. This completes the proof 
that the BFKL result we obtained is consistent 
with infrared factorisation.

\section{Conclusions} 
\label{conclusion}

We considered the even signature component of 
two-to-two parton scattering amplitudes in the 
high-energy limit. This amplitude is dominated by 
the $t$-channel exchange of a state consisting of 
two Reggeized gluons, corresponding to the simplest 
example of a Regge cut in QCD. The amplitude can 
be evaluated in QCD perturbation theory by iteratively 
solving the BFKL equation. Each order in perturbation 
theory corresponds to one additional rung in the BFKL 
ladder, building up a tower of so-called next-to-leading 
logarithms, ${\cal O}(\alpha_s^{\ell} L^{\ell-1})$. 
Although the BFKL Hamiltonian has been diagonalised 
in many cases \cite{Lipatov:1985uk}, the 
dimensionally-regulated Hamiltonian relevant
for partonic amplitudes has remained more difficult 
to handle.

Our first observation was that the wavefunction 
describing the two Reggeized gluons remains 
finite through BFKL evolution for any number of 
rungs, while the corresponding amplitude develops 
infrared singularities due to the soft limit of the 
wavefunction. We further observed that the 
evolution of a state in which one of the two 
Reggeized gluons is much softer than the other, 
$k\ll p-k$, yields again a similar state. In other 
words, the soft approximation is consistent with
BFKL evolution, and as a consequence, one 
can systematically solve the equation to any 
loop order within this approximation. 
We found that the soft approximation leads 
to a major simplification, where all integrals 
reduce to products of bubbles, and the 
wavefunction at any given order is simply 
a polynomial of that order in $\left(p^2/k^2\right)^{\eps}$. 
This eventually allowed us to determine the 
singularities of the amplitude in a closed form 
to any order, as given in \eqn{eq:mredregge}. 

At the next step we contrasted the singularity 
structure we obtained though BFKL evolution 
with the known exponentiation properties of 
infrared singularities. As expected, we found 
that the two are consistent, and this provides 
a highly non-trivial check of the calculation. 
The leading singularity at each order, 
${\cal O}(\alpha_s^{\ell}L^{\ell-1}/\epsilon^{\ell})$, 
is simply related to the one-loop soft anomalous 
dimension, and has a colour structure proportional 
to $(C_A-\Tt)^{\ell-1}$. New singularities, with fewer 
powers of $1/\eps$ and different colour structures, 
appear starting from four loop. These correspond 
to new terms in the imaginary part of the soft 
anomalous dimension, \eqn{eq:gamma8}. We 
were thus able to determine the soft anomalous 
dimension at next-to-leading logarithmic accuracy 
in the high-energy limit to all orders. These results 
also provide a valuable input for determining the 
structure of long-distance singularities for general 
kinematics using a bootstrap approach, as done 
at the three-loop order in ref.~\cite{Almelid:2017qju}. 

We point out that the $\ell$-loop coefficient of the 
soft anomalous dimension we computed is a linear 
combination of zeta values of weight $(\ell-1)$, 
which coincides with the maximal (transcendental) 
weight. This is not surprising given that these 
corrections are independent of the matter content 
nor the amount of supersymmetry of the theory, 
and are thus common for example to QCD and 
${\cal N}=4$ super Yang-Mills. We further showed 
that these corrections to the soft anomalous dimension 
can be resummed, as in \eqn{GammaNLL3}, into an 
entire function of $x=L\alpha_s/\pi$. 
Remarkably, this gives us means
to determine the asymptotic high-energy behaviour 
of this anomalous dimension, corresponding to 
$x\gg1$, a regime which is usually inaccessible to 
perturbation theory. We find that at large $x$ the 
imaginary part of the anomalous dimension in the 
Regge limit, in any colour representation, becomes 
an oscillating function with an exponentially 
growing amplitude.

While our analysis in this paper was focused on 
infrared singularities, for which the soft approximation 
is sufficient, the formulation of the evolution in \eqn{Him} 
along with the observation that the wavefunction is finite, 
pave the way to determining the wavefunction beyond the 
soft approximation, thus evaluating the finite contributions 
to Regge-cut of two-to-two amplitudes.  It would also be 
interesting to extend the present analysis to the next order, 
using the known next-to-leading order Hamiltonian; again 
we expect that a suitable wavefunction will remain finite 
to all orders, facilitating a direct determination of the 
infrared singularities.

%%%%%%%%%%%%%%%%%%%%%%%%%%%%%%%%%%%%%%%%%
\vspace{20pt}
\acknowledgments

We would like to thank J.M.~Smillie for useful 
discussions in the early stages of this project. 
SCH's research is supported by the National Science 
and Engineering Council of Canada, and was supported 
in its early stage by the Danish National Research 
Foundation (DNRF91). EG's  research is supported by 
the STFC Consolidated Grant ``Particle Physics at the 
Higgs Centre.'' LV's research is supported by the People 
Programme (Marie Curie Actions) of the European Union's 
Horizon 2020 Framework Programme H2020-MSCA-IF-2014 
under REA grant No.~656463 -- ``Soft Gluons''. SCH thanks 
the Higgs Centre for Theoretical Physics for hospitality during 
part of this work. This research was conducted in part at the 
CERN summer institute ``LHC and the Standard Model: 
Physics and Tools'' and at the workshops ``Automated, 
Resummed and Effective: Precision Computations for 
the LHC and Beyond'' and ``Mathematics and Physics 
of Scattering Amplitudes''  at the Munich Institute for 
Astro- and Particle Physics (MIAPP) of the DFG 
cluster of excellence ``Origin and Structure of the 
Universe''.

%%%%%%%%%%%%%%%%%%%%%%%%%%%%%%%%%%%%%%%%%

\appendix

\section{The even amplitude at NLL accuracy within the 
shockwave formalism}\label{Review}

In this appendix we briefly review how \eqn{ReducedAmpNLL2}
can be derived within the shockwave formalism 
refs.~\cite{Caron-Huot:2013fea,Caron-Huot:2017fxr}.
Amplitudes in the high-energy limit 
are calculated as expectation values of null Wilson lines:
\be
U(z_\perp) =  \mathcal{P}\exp\left[ig_s\int_{-\infty}^{+\infty}\,\Dd x^+\,
A_+^a(x^+,x^-{=\, }0,z_\perp)\, T^a\right].
\ee
The latter follows the path of colliding partons from the
projectile or target (with $x^+$ and $x^-$ interchanged), and
are labelled by transverse coordinates $z_\perp$ (below 
we shall omit the subscript $\perp$ for lighter notation).
The full transverse structure needs to be retained, 
because the high-energy limit is taken with fixed 
momentum transfer.  Importantly, the number 
of Wilson lines cannot be held fixed, because the projectile 
and target contain an arbitrary number of virtual partons.
However, in perturbation theory, the unitary matrices $U(z)$ 
are close to the identity and can therefore be usefully 
parametrised by a field $W$:
\be \label{Uparam}
U(z) = e^{ig_s\,T^aW^a(z)}\,.
\ee
Physically, the colour-adjoint field $W^a$, which 
is propagating in the transverse space, is interpreted 
as source for a BFKL Reggeized gluon~\cite{Caron-Huot:2013fea}.
At weak coupling a generic projectile is thus formed 
by a superposition of $W$ states. Up to NLL
accuracy one needs to consider up to two Reggeons.
In this approximation, a projectile, created with four-momentum $p_1$ 
and absorbed with $p_4$, is parameterised in 
momentum space as 
\bea\label{OPE} 
\ket{\psi_i}  &\equiv& \frac{Z_i^{-1}}{2p_1^+} a_i(p_4) a^\dagger_i(p_1)\ket{0} 
\,=\, \ket{\psi_{i,1}}+\ket{\psi_{i,2}} +\ldots\,,
\eea
where the ellipses stand for wavefunction components with 
three or more Reggeized gluons, which are not relevant at 
NLL accuracy. We next note that states with an even (odd) 
number of Reggeized gluons have an even (odd) signature, 
so
\begin{subequations}
\begin{align}
\ket{\psi_{i,1}}&=\ket{\psi_{i,1}^{(-)}}= i g_s \, D_i^{(1)}(p) \T_i^a\,  W^a(p) \label{psi1} \\
\ket{\psi_{i,2}}&=\ket{\psi_{i,2}^{(+)}}= - \frac{g_s^2}{2} \T_i^a\T_i^b \label{psi2}
\int \frac{\Dd^{2{-}2\eps} q}{(2\pi)^{2-2\eps}} \,\Omega^{(0)}(p,q)\, W^a(q)W^b(p{-}q),
\end{align}
\end{subequations}
where $D_i^{(1)}(p)$ is an impact factor which 
parameterises the dependence of the coefficient on 
the (transverse) momentum transfer $p=p_4-p_1$ 
with $p^2 =- t$. At the leading order, there is only 
one Wilson line $U(z)$ following the original parton, and 
the two-Reggeon wavefunction is obtained simply 
by expanding \eqn{Uparam}, which gives, as in 
the main text:
\be
 \Omega^{(0)}(p,q) =1.
\ee
The null Wilson lines acquire energy dependence 
through rapidity divergences, which must be regulated, 
leading to the Balitsky-JIMWLK  rapidity evolution equation:
\be \label{rapidity_evolution}
\frac{d}{d\eta}\,\ket{\psi_i} = H\, \ket{\psi_i}\,.
\ee
The scattering amplitude can be obtained by computing 
the overlap between $\bra{\psi_{j}} $ and $\ket{\psi_i}$,
after evolving them to common rapidity, where the 
overlap is defined as the vacuum expectation value
of left-moving and right-moving $W$-fields. In terms 
of the reduced amplitude defined in \eqn{Mreduced} 
one has  
\be \label{reduced_amp_from_inner_product}
\frac{i}{2s} \Mreduced_{ij\to ij} = 
\bra{\psi_{j}}e^{\Hhat\Log}\ket{\psi_i},
\qquad \Hhat \equiv H-\T_t^2 \, \alpha_g(t).
\ee
Evolution at the desired accuracy is obtained 
by simply considering the Hamiltonian at leading 
order in $g_s^2$ in terms of $W$ fields, which, to 
this order, is diagonal:
\bea \label{Hamiltonian_schematic_form}
\hat H  \left( 
\begin{array}{c}
  W      \\  WW   
\end{array}
\right) &\equiv&
\left(
\begin{array}{ccc}
 \hat H_{1{\to}1} & 0  & \\
 0 & \hat H_{2{\to}2}  & \\
\end{array}
\right)\left(
\begin{array}{c}
  W      \\  WW   \end{array}
\right) + {\cal O}(g_s^4).
\eea
Since the signature odd and even 
sectors are orthogonal and closed under the 
action of $\Hhat$ (as a consequence of the 
signature symmetry), their contributions to 
the amplitude at NLL separate:
\bea \label{Regge-odd-Even-Amplitude} \nn
\frac{i}{2s}\Mreduced^{\rm NLL}_{ij\to ij} &=&
\frac{i}{2s}\left( 
\Mreduced^{(-),\rm NLL}_{ij\to ij}+
\Mreduced^{(+),\rm NLL}_{ij\to ij} \right) \\[0.2cm]
&\equiv& \bra{\psi^{(-)}_{j,1}}e^{\Hhat\Log}\ket{\psi^{(-)}_{i,1}}^{\NLO}
+ \bra{\psi^{(+)}_{j,2}}e^{\Hhat\Log}\ket{\psi^{(+)}_{i,2}}^{\LO},
\eea
where ``LO'' and ``NLO'' means that all ingredients 
are needed respectively to leading and next-to-leading
nonvanishing order. In this paper we  focus on the 
even amplitude, representing the exchange of a pair of 
Reggeons,  corresponding to the second term in 
eq.~(\ref{Regge-odd-Even-Amplitude}). It is then 
convenient to compute the inner product in 
\eqn{reduced_amp_from_inner_product} by 
first evolving the wavefunction:
\be
e^{\Hhat_{2\to 2}\Log}\ket{\psi_{i,2}^{(+)}}= -\frac{g_s^2}{2} \T_i^a\T_i^b \label{psi2a}
\sum_{\ell=0}^\infty \frac{1}{\ell!}\left(\frac{\as B_0(\eps)L}{\pi}\right)^\ell
\int \frac{\Dd^{2{-}2\eps} q}{(2\pi)^{2-2\eps}} \,\Omega^{(\ell)}(p,q)\, W^a(q)W^b(p{-}q)\,.
\ee
As displayed in \eqn{Hdef},
the wavefunctions $\Omega^{(\ell)}$ may then be 
obtained iteratively by applying the Hamiltonian 
$\hat H_{2{\to}2}$. This Hamiltonian was discussed at length 
in terms of Wilson lines in ref.~\cite{Caron-Huot:2017fxr}, 
to which we refer for further details ($-H_{k\to k}$ 
is given in eq. (3.13) there; note the overall minus sign 
between our conventions). Acting with $\hat H_{2{\to}2}$ on the states 
in \eqn{psi2a}, reproduces precisely the leading order BFKL 
Hamiltonian recorded in the main text. Finally, computing 
the overlap with the target state $\bra{\psi^{(+)}_{j,2}}$
produces the integral which closes the ladder in 
\eqn{ReducedAmpNLL2}.

\section{Proof of the all-order amplitude}
\label{AppB}

In this appendix we show that the singular terms 
in \eqn{MellReggeSoft-All-B} are equal to those 
in \eqn{MellReggeSoft-ResB}. We start by 
noticing that the statement is equivalent to 
\be 
\sum_{n=1}^{\ell} 
(-1)^{n+1} \, \binom{\ell}{n} \prod_{m=0}^{n-2}\bigg[ 1 - \hbn{m}(\eps) \frac{2 C_A -\Tt}{C_A -\Tt} \bigg]
- \left( 1- \hbn{-1}(\eps) \frac{2C_A-\Tt}{C_A -\Tt} \right)^{-1} = {\cal O}(\eps^{\ell}).
\ee
Multiplying both sides of this equality by 
$ \left( 1- \hbn{-1}(\eps) \frac{2C_A-\Tt}{C_A -\Tt} \right) = 1+{\cal O}(\eps^3)$
we get 
\be 
\sum_{n=1}^{\ell} 
(-1)^{n+1} \, \binom{\ell}{n} \prod_{m=0}^{n-2}\bigg[ 1 - \hbn{m}(\eps) \frac{2 C_A -\Tt}{C_A -\Tt} \bigg]
\left( 1- \hbn{-1}(\eps) \frac{2C_A-\Tt}{C_A -\Tt} \right) - 1 = {\cal O}(\eps^{\ell}).
\ee
The additional factor multiplying the 
sum on the l.h.s.\ can be incorporated into the 
product. Similarly, the $-1$ on the l.h.s.\ 
can be included in the sum. We obtain
\be \label{almost-proven}
\sum_{n=0}^{\ell} 
(-1)^{n+1} \, \binom{\ell}{n} \prod_{m=-1}^{n-2}\bigg[ 1 - \hbn{m}(\eps) \frac{2 C_A -\Tt}{C_A -\Tt} \bigg]
= {\cal O}(\eps^{\ell}).
\ee
At this point, we realise that the 
structure of the sum and product is 
strikingly similar to that appearing in the 
target-averaged wavefunction in \eqn{Well-1-ansatz}. 
In that case, finitness of the $\ell$-loop wavefunction 
implies
\be \label{Well-eps-ell}
  \sum_{n=0}^{\ell} (-1)^n n^q \binom{\ell}{n} \prod_{m=0}^{n-1} 
  \left[1 - \hbn{m}(\eps) \frac{2\CA-\Tt}{\CA-\Tt}\right]
  = \ord \left(\eps^{\ell-q}\right) \quad \text{with} \quad q 
  = 0,1,2,\dots
\ee
which is obtained by expanding 
$(p^2/k^2)^{n\eps}$ around small 
$\eps$ inside the sum. Next, we 
bring the product in~\eqn{almost-proven} to the same 
form as in~\eqn{Well-eps-ell}, obtaining
\bea \label{proven} \nn
\sum_{n=0}^{\ell} 
(-1)^{n+1} \, \binom{\ell}{n} 
\left(1 - \hbn{-1}(\eps) \frac{2 C_A -\Tt}{C_A -\Tt}\right)
\left(1 - \hbn{n-1}(\eps) \frac{2 C_A -\Tt}{C_A -\Tt}\right)^{-1} && \\
&&\hspace{-6.0cm}\times\, \prod_{m=0}^{n-1} 
\bigg[ 1 - \hbn{m}(\eps) \frac{2 C_A -\Tt}{C_A -\Tt} \bigg]
= {\cal O}(\eps^{\ell}).
\eea
The extracted factor
\begin{multline} \label{eq:addfac}
\left(1 - \hbn{-1}(\eps) \frac{2 C_A -\Tt}{C_A -\Tt}\right)
\left(1 - \hbn{n-1}(\eps) \frac{2 C_A -\Tt}{C_A -\Tt}\right)^{-1} \\
= 1 + \frac{2C_A-\Tt}{C_A-\Tt} \left[2 \eps (n\eps)^2 \zeta_3 
+ 3 \eps^2 (n\eps)^2 \zeta_4 + \left(4\eps^3(n\eps)^2 
+ 2 \eps (n\eps)^4\right) \zeta_5\right] + \ord(\eps^6)
\end{multline}
is a function of $\eps$ and $\delta \equiv n\eps$, 
cf.\ eqs.~\eqref{bubbleGeneral2} and 
\eqref{bubblehat}, which are both small. In 
other words, the (double) expansion of 
\eqn{eq:addfac} in $\eps$ and $\delta$ 
around 0 contains only terms for which 
the power of $\eps$ is equal or greater 
than the power of $n$. This, then, together 
with \eqn{Well-eps-ell}, proves 
\eqn{almost-proven} and thus the 
conjectured amplitude 
\eqref{MellReggeSoft-All-B}.

\bibliography{Draft-v16}
\bibliographystyle{JHEP}

\end{document}